\pdfoutput=1
\documentclass[12pt,a4paper]{article}
\usepackage{amsmath,amssymb,type1cm,booktabs,bm,mymacros}
\usepackage[pdftex]{graphicx}
\allowdisplaybreaks[1]
\numberwithin{equation}{section}

\begin{document}
\begin{titlepage}
\begin{flushright}
Cavendish-HEP-09/21 \\
DAMTP-2009-71 \\
\end{flushright}

\vspace{1ex}

\begin{center}

{\LARGE\bf Gauge-mediated supersymmetry breaking with generalized
 messenger sector at LHC}
 
\vspace{3ex}

{\large $^{(a)}$Hidetoshi Kawase\footnote{e-mail: hkawase@eken.phys.nagoya-u.ac.jp},
 $^{(a)}$Nobuhiro Maekawa\footnote{e-mail: maekawa@eken.phys.nagoya-u.ac.jp},
 $^{(b)}$Kazuki Sakurai\footnote{e-mail: sakurai@hep.phy.cam.ac.uk} 
}

\vspace{4ex}
{\it $^a$Department of Physics, Nagoya University, Nagoya 464-8602, Japan} \\
\vspace{2ex}
{\it $^b$DAMTP, Wilberforce Road, Cambridge, CB3 0WA, UK \\
 Cavendish Laboratory, JJ Thomson Avenue, Cambridge, CB3 0HE, UK} \\

\vspace{6ex}

\end{center}

\begin{abstract}
 We consider the generalized gauge mediated supersymmetry breaking
(GMSB) models with the messenger fields which do not form the complete
multiplets of $SU(5)$ GUT symmetry.
Such a situation may happen in the anomalous $U(1)$ GUT scenario
because the mass spectrum of the superheavy particle does not
respect $SU(5)$ GUT symmetry, although the success of the gauge
coupling unification can be explained.
In this paper, we assume that one pair of the messenger fields gives
the dominant contribution, and the LHC signature for the two possible
messengers, $X + \bar{X}$ and $Q + \bar{Q}$, are examined.
We investigate the possibility to measure
the deviation from the usual GUT relation of the gaugino masses which is
one of the most important features of these scenarios.

\end{abstract}
\end{titlepage}
\section{Introduction}
The minimal supersymmetric (SUSY) standard model (MSSM) is one of the most 
promising candidates for the model beyond the standard model (SM)
\cite{Nilles:1983ge,Haber:1984rc,Martin:1997ns}.
Unfortunately, the MSSM has more than 100 parameters concerning on the 
SUSY breaking, and the signatures in the large hadron collider (LHC) are 
strongly dependent on the parameters, especially, the mass spectrum of 
SUSY particles
\cite{Aad:2009wy,Ball:2007zza,Hinchliffe:1996iu,Hinchliffe:1998ys,
Bachacou:1999zb,Allanach:2000kt,Gjelsten:2004ki}.
Therefore, it is important to study various possibilities
for the SUSY breaking parameters and the LHC signatures before the LHC 
starts to present the data. In this paper, we examine the generalized gauge
mediated SUSY breaking (GMSB) scenario
\cite{Martin:1996zb,Meade:2008wd,Marques:2009yu}
in which the messenger fields do not form the complete grand unified
theory (GUT) multiplets.

In the usual GMSB scenario \cite{Dine:1993yw,Dine:1994vc},
the messenger fields are adopted as complete multiplets under $SU(5)$.
This is mainly because the messenger fields which do not respect
$SU(5)$ generically spoil the success of the gauge coupling
unification in the MSSM.
However, it has been understood that the GUT with anomalous $U(1)$ gauge
symmetry \cite{Maekawa:2001uk,Maekawa:2002bk,Maekawa:2003bb}
can naturally explain the
success of the gauge coupling unification
in the MSSM although the superheavy particles do not respect the $SU(5)$
symmetry \cite{Maekawa:2001vt,Maekawa:2002mx}.
Since some of the superheavy particles can play the same role as the messenger
fields, it is important to study the generalized GMSB
scenario with the messenger fields which do not form the complete
multiplets of $SU(5)$.

One of the most interesting facts in this scenario is that the GUT relation
for the gaugino masses are spoiled 
although the anomalous $U(1)$ GUT has the GUT 
gauge symmetry at the GUT scale.
Actually, if we assume that the SUSY is broken by some mechanism and
the resulting superpotential has a form
\begin{equation}
W = m_{\Phi}\Phi\bar{\Phi} + \theta^2F_{\Phi}\Phi\bar{\Phi},
\end{equation}
where $\theta^2$ is a superspace coordinate, 
the masses of gauginos and sfermions at the scale $m_{\Phi}$
are generated by one and two loop effects
of the messenger fields $\Phi$ and $\bar\Phi$ as \cite{Dine:1993yw,Dine:1994vc}
\begin{equation}
M_a = n_a\left(\frac{\alpha_a}{4\pi}\right)\frac{F_\Phi}{m_\Phi}
+ \mathcal{O}\left(\frac{F_\Phi^3}{m_\Phi^5}\right),
\end{equation}
\begin{equation}
m_{\tilde f}^2 = \sum_{a=1}^3n_aC_a^{\tilde f}
 \left(\frac{\alpha_a}{4\pi}\right)^2\frac{F_\Phi^2}{m_\Phi^2}
 + \mathcal{O}\left(\frac{F_\Phi^4}{m_\Phi^6}\right),
\end{equation}
respectively. Here $n_a$ ($a=1,2,3$) is the Dynkin index whose
normalization is chosen to be $n_a=1$ for $\bm{5} +\bar{\bm{5}}$ of $SU(5)$,
and $C_a^{\tilde f}$ is the quadratic Casimir invariant of sfermions.
$C_a^{\tilde f}$ are in a normalization where
$C_1^{\tilde f}=3/5\cdot Y^2$ for sfermions with hypercharge $Y$,
$C_2^{\tilde f}=3/4$ for $SU(2)_L$ doublets and $C_3^{\tilde f}=4/3$
for $SU(3)_C$ triplets.
$n_a$ for various messenger fields are given in Table \ref{tab:n}.
The generalized messenger scenario has a lot of possibilities in general
\cite{Martin:1996zb,Meade:2008wd}.
Here, just for simplicity, we assume that one of the messenger fields 
in Table \ref{tab:n} dominates.
Then only two possibilities, $X + \bar{X}$ and $Q + \bar{Q}$,
can give the non-vanishing masses to all the gauginos. In this paper, 
we examine these two possibilities.

\begin{table}
 \label{tab:n}
 \begin{center}
 \begin{tabular}{clccc} \toprule\hline \\[-12pt]
  & $\quad (SU(3)_C,SU(2)_L)_{U(1)_Y}$ & $\quad n_1 \quad$
  & $\quad n_2 \quad$ & $\quad n_3 \quad$ \\[3pt] \hline \\[-12pt]
  $\quad Q + \bar{Q} \quad$
  & $\quad (\bm{3},\bm{2})_{1/6}+(\bar{\bm{3}},\bm{2})_{-1/6}$
      & $1/5$ & $3$ & $2$ \\[3pt]
  $U + \bar{U}$ & $\quad (\bm{3},\bm{1})_{2/3}+(\bar{\bm{3}},\bm{1})_{-2/3}$
      & $8/5$ & $0$ & $1$ \\[3pt]
  $D + \bar{D}$ & $\quad (\bm{3},\bm{1})_{-1/3}+(\bar{\bm{3}},\bm{1})_{1/3}$
      & $2/5$ & $0$ & $1$ \\[3pt]
  $L + \bar{L}$ & $\quad (\bm{1},\bm{2})_{-1/2}+(\bm{1},\bm{2})_{1/2}$
      & $3/5$ & $1$ & $0$ \\[3pt]
  $E + \bar{E}$ & $\quad (\bm{1},\bm{1})_{-1}+(\bm{1},\bm{1})_{1}$
      & $6/5$ & $0$ & $0$ \\[3pt]
  $G$ & $\quad (\bm{8},\bm{1})_0$ & $0$ & $0$ & $3$\\[3pt]
  $W$ & $\quad (\bm{1},\bm{3})_0$ & $0$ & $2$ & $0$\\[3pt]  
  $X + \bar{X}$ & $\quad (\bm{3},\bm{2})_{-5/6}+(\bar{\bm{3}},\bm{2})_{5/6}$
      & $5$ & $3$ & $2$ \\[-12pt] \\ \hline \bottomrule
 \end{tabular}
 \caption{The Dynkin indices $n_a$ $(a = 1,2,3)$ for various messenger
 fields.}   
 \end{center}
\end{table}

\section{Overview of the mass spectrum}
As discussed in the introduction,
the GUT relation for the gaugino masses is generally spoiled in the generalized
GMSB scenario.
Since $M_a\alpha_a^{-1}$ is one-loop renormalization group invariant,
the gaugino masses satisfy the relation
\begin{equation}
 M_1 : M_2 : M_3 \sim n_1\alpha_1 : n_2\alpha_2 : n_3\alpha_3
\end{equation}
at any renormalization scale.
Therefore, if we consider a model with messengers which respect $SU(5)$
symmetry, the ratio of gaugino masses at weak scale is given as
\begin{equation}
 \label{eq:gutrel}
 M_1(m_Z) : M_2(m_Z) : M_3(m_Z) \sim 1 : 2 : 6.
\end{equation}
The relation \eqref{eq:gutrel} is often called the GUT relation of the gaugino
masses.
However, this relation is spoiled if $n_1=n_2=n_3$ is not satisfied as in 
the generalized GMSB scenario. 
In the followings, we examine the spectra of models with $X + \bar{X}$
or $Q + \bar{Q}$ messengers as a specific example of such scenarios.
The parameters for the SUSY breaking sector are 
$\Lambda_{\Phi} \equiv F_{\Phi}/m_{\Phi}$,
$m_{\Phi}$, $\tan\beta$ (the ratio of the VEVs of up-type Higgs and 
the down-type Higgs), and $\mathrm{sgn}\,(\mu)$ (the sign of the SUSY Higgs
mass).
The masses of gauginos and sfermions, and the 
scalar trilinear couplings $A$ at the mass 
scale of the messenger fields are given as
\begin{equation}
 M_a(m_{\Phi}) \simeq
  n_a\left(\frac{\alpha_a}{4\pi}\right)\Lambda_{\Phi},
  \qquad
  m_{\tilde{f}}^2(m_{\Phi}) \simeq
  \sum_{a=1}^3n_aC_a^{\tilde{f}}
  \left(\frac{\alpha_a}{4\pi}\right)^2\Lambda_{\Phi}^2,
  \qquad
  A(m_{\Phi}) \simeq 0.
 \label{masses}
\end{equation}
We use the renormalization group equations (RGEs) to obtain these
parameters at the weak scale.
For implementing the numerical calculation, 
we use {\sf SOFTSUSY 2.0.18} \cite{Allanach:2001kg}
with appropriate modification according to our purpose.
In our calculation, we assume that the contributions from messenger
fields other than the selected one are relatively small
and can be neglected entirely.
To specify the SUSY Higgs mass $\mu$ and the Higgs mixing parameter
$b$ we use the relations
\begin{equation}
 b = \frac{1}{2}(m_{H_u}^2 + m_{H_d}^2 + 2|\mu|^2)\sin 2\beta
\end{equation}
and
\begin{equation}
 |\mu|^2 =
  \frac{m_{H_d}^2 - m_{H_u}^2\tan^2\beta}{\tan^2\beta - 1}
  - \frac{m_Z^2}{2}.
\end{equation}
where $m_{H_u}^2$ and $m_{H_d}^2$ are the SUSY breaking Higgs
mass parameters at the weak scale which are calculated by
the RGEs with the boundary values given by \eqref{masses}
at the scale $m_{\Phi}$.

First, we consider the scenario with $X + \bar{X}$ messenger.
In this scenario, the Dynkin indices are given by
\begin{equation}
 n_1 = 5,\qquad n_2 = 3,\qquad n_3 = 2
\end{equation}
as shown in Table \ref{tab:n},
so gaugino masses at the weak scale satisfy the relation
\begin{equation}
 \label{eq:xrel}
 M_1(m_Z) : M_2(m_Z) : M_3(m_Z) \sim 5 : 6 : 12
\end{equation}
at one-loop order.
Since the relation \eqref{eq:xrel} is not affected so much by the specific
choice of the parameters such as the mass scale of messenger
particles, we can use this relation to distinguish this model from
others.
The relation \eqref{eq:xrel} indicates that the hierarchy among
the gaugino masses becomes milder than the usual GUT relation.
One of the most interesting features is that the mass splitting between
the bino and wino is especially small. Therefore, to check this feature
is one of the promising ways to test this scenario.

\begin{figure}
 \begin{center}
   \includegraphics[scale=1.0]{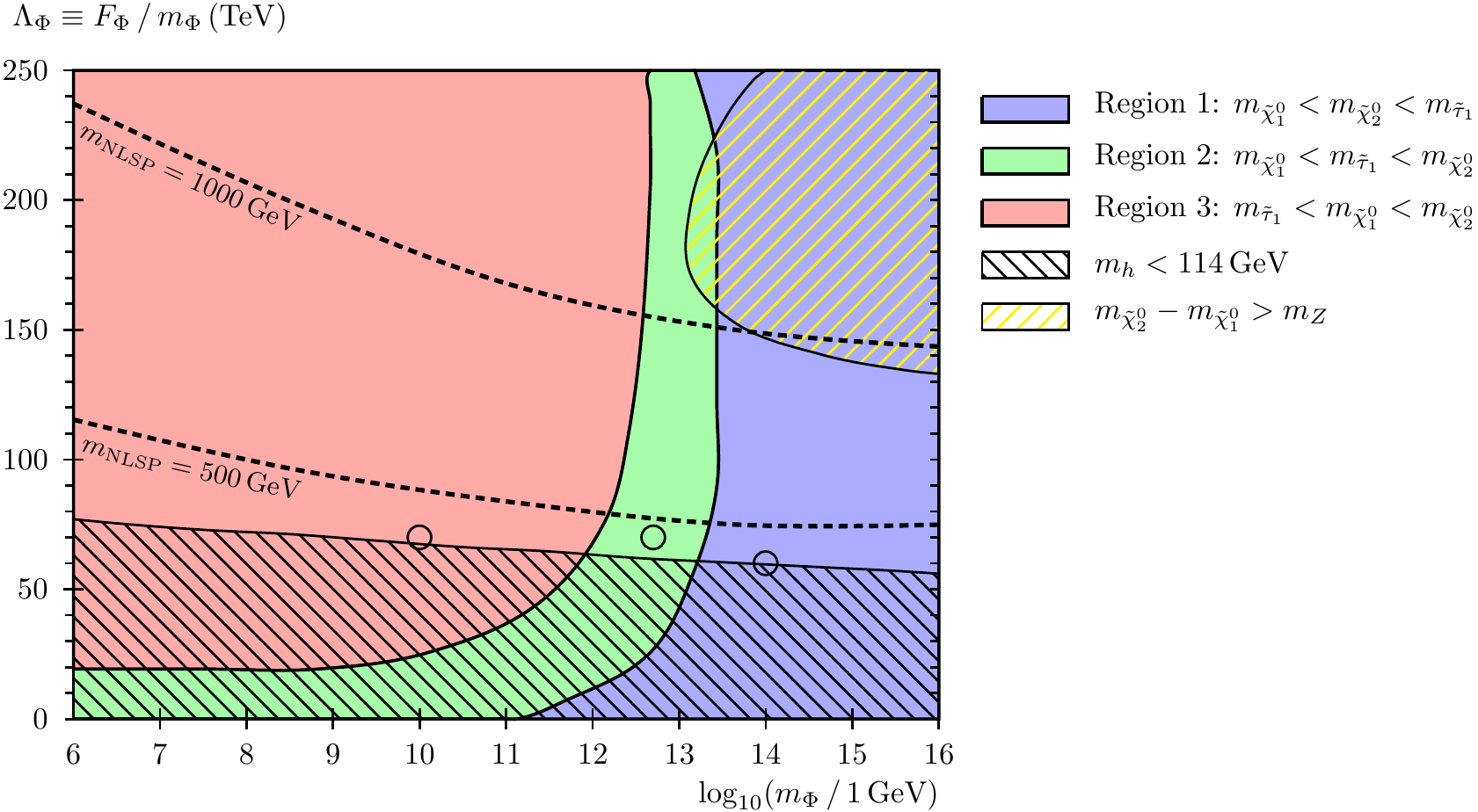}
 \caption{Allowed parameter region for $X + \bar{X}$ messenger scenario.
  We set $\tan\beta = 10$ and $\mathrm{sgn}\,(\mu) = +1$.
  The circles correspond to the model points selected to analyze the
  LHC signature.}
  \label{fig:regionx}
 \end{center}
\end{figure}

Figure \ref{fig:regionx} represents the parameter space in this
scenario. Here we set $\tan\beta = 10$ and $\mathrm{sgn}\,(\mu) = +1$,
and we assume that the lightest SUSY particle (LSP) is the gravitino.
The experimental bound for this scenario comes mainly from the LEP2
bound on the lightest Higgs mass $m_h > 114.4\,\mathrm{GeV}$
\cite{Amsler:2008zzb}.
The mass of Higgs is, however, largely dependent on the mass of
top quark, so there remains a large uncertainty concerning this bound.
We set $m_t = 175\,\mathrm{GeV}$ for our calculation.
As shown in Figure \ref{fig:regionx}, there are 
three parameter regions, corresponding to
\begin{enumerate}
 \item $m_{\tilde{\chi}_1^0} < m_{\tilde{\chi}_2^0} < m_{\tilde{\tau}_1}$
 \item $m_{\tilde{\chi}_1^0} < m_{\tilde{\tau}_1} < m_{\tilde{\chi}_2^0}$
 \item $m_{\tilde{\tau}_1} < m_{\tilde{\chi}_1^0} < m_{\tilde{\chi}_2^0}$.
\end{enumerate}
When the $\Lambda_{\Phi}$ is comparatively small, the 
higgsinos are relatively heavy compared
with the gauginos, and therefore we can roughly identify the 
lightest neutralino
$\tilde{\chi}_1^0$ with the  bino and the second lightest neutralino
$\tilde{\chi}_2^0$ with the wino.
Therefore, the bino-like neutralino $\tilde{\chi}_1^0$ becomes
the next to LSP (NLSP)
 in the regions 1 and 2, and the stau $\tilde{\tau}_1$ becomes the NLSP 
 in the region 3.

\begin{figure}
 \begin{center}
   \includegraphics[scale=1.0]{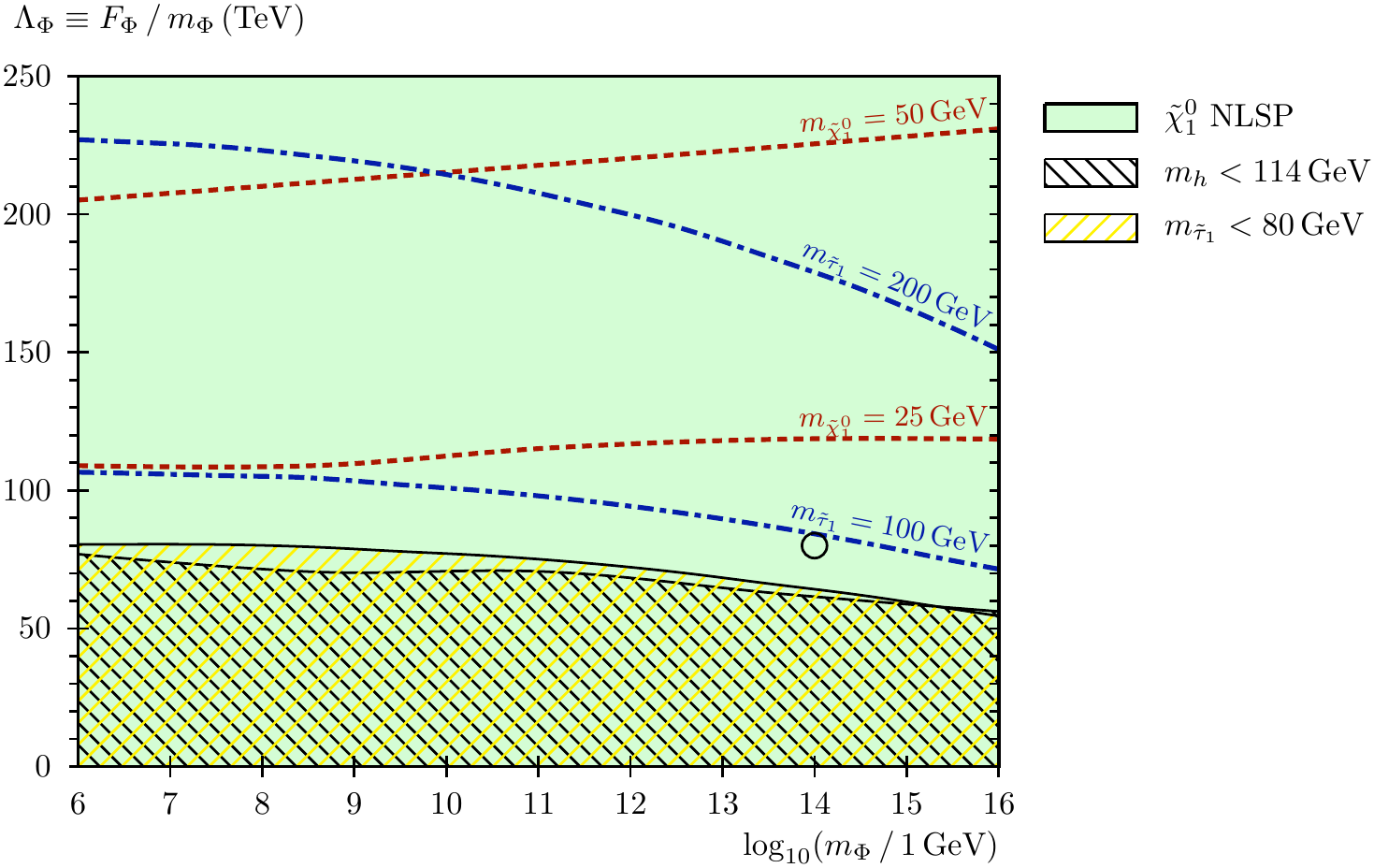}
 \caption{Allowed parameter region for $Q + \bar{Q}$ messenger scenario.
  We set $\tan\beta = 10$ and $\mathrm{sgn}\,(\mu) = +1$.
  The circle corresponds to the model point selected to analyze the
  LHC signature.}
 \label{fig:regionq}  
 \end{center}
\end{figure}
Another candidate of messenger fields in our scenario is the fields
with quantum number of $Q + \bar{Q}$.
In this scenario, the Dynkin indices are given as
\begin{equation}
 n_1 = \frac{1}{5},\qquad n_2 = 3,\qquad n_3 = 2
\end{equation}
and gaugino masses satisfy the relation
\begin{equation}
 M_1(m_Z) : M_2(m_Z) : M_3(m_Z) \sim \frac{1}{5} : 6 : 12
\end{equation}
at one-loop order.
As can  be seen from this relation, $Q + \bar{Q}$ messenger
scenario gives rather small masses to sparticles which do not have
quantum numbers of $SU(2)_L$ and $SU(3)_C$ compared with the other sparticles.
The allowed parameter space is presented in Figure \ref{fig:regionq}.
In the whole allowed region,
$m_{\tilde{\chi}_1^0} < m_{\tilde{\tau}_1} < m_{\tilde{\chi}_2^0}$
is satisfied as in the region 2 of $X + \bar{X}$ messenger scenario,
and therefore the NLSP is the bino-like neutralino $\tilde{\chi}_1^0$.

Note that  there are almost no bounds for the mass of the lightest
neutralino $\tilde{\chi}_1^0$ if $\tilde{\chi}_1^0$ is the pure bino
and does not decay inside the detector.
The GUT relation is essential to obtain the bound
$m_{\tilde{\chi}_1^0} > 46\,\mathrm{GeV}$ given by the particle data group
\cite{Amsler:2008zzb} and the constraint from
the invisible decay of $Z$ is useless because the decay width
of $Z \to \tilde{\chi}_1^0\,\tilde{\chi}_1^0$ is quite small for
the bino-like $\tilde{\chi}_1^0$ \cite{Dreiner:2009ic}.
Therefore the constraint for the mass of right-handed stau
$\tilde{\tau}_1$, $m_{\tilde{\tau}} > 81.9\,\mathrm{GeV}$ is important
in this scenario which is shown in Figure \ref{fig:regionq}.

\section{LHC signature}
In this section, we investigate the LHC signatures
of these scenarios.
For this purpose, we use {\sf ISAJET 7.79} \cite{Paige:2003mg}
to calculate the decay width
of sparticles and {\sf HERWIG 6.510}
\cite{Corcella:2000bw,Corcella:2002jc}
to generate the sparticle production
events by Monte-Carlo simulation.
And for the detector simulation, we use {\sf AcerDET 1.0}
\cite{RichterWas:2002ch} as a
fast simulation of the search at the LHC.
We examine the LHC signatures for $\sqrt{s} = 14\,\mathrm{TeV}$ for the
whole analysis in this paper.

\begin{table}
 \begin{center}
  \begin{tabular}{cccccccc} \toprule \hline
   & $\Lambda_{\text{mess}}$ & $m_{\text{mess}}$ & $\tan\beta$
   & $\mathrm{sgn}\,(\mu)$ & $n_1$ & $n_2$ & $n_3$ \\
   & (TeV) & (GeV)
   \\ \hline
   \quad Case 1:
   $m_{\tilde{\chi}_1^0} < m_{\tilde{\chi}_2^0} < m_{\tilde{l}_R}$\quad &
       $60$ & $1.0 \times 10^{14}$ & $10$ & $+$ & $5$ & $3$ & $2$ \\   
   \quad Case 2:
   $m_{\tilde{\chi}_1^0} < m_{\tilde{l}_R} < m_{\tilde{\chi}_2^0}$\quad &
       $70$ & $5.0 \times 10^{12}$ & $10$ & $+$ & $5$ & $3$ & $2$ \\
   \quad Case 3:
   $m_{\tilde{l}_R} < m_{\tilde{\chi}_1^0} < m_{\tilde{\chi}_2^0}$\quad &
       $70$ & $1.0 \times 10^{10}$ & $10$ & $+$ & $5$ & $3$ & $2$
			       \\ \hline \bottomrule
  \end{tabular}
 \end{center}
 \begin{center}
  \begin{tabular}{cc}
   \begin{tabular}{crrr} \toprule \hline
    & Case 1 & Case 2 & Case 3 \\ \hline
    $\quad \tilde{g} \quad$ & 910 & 1049 & 1054 \\
    $\tilde{u}_L$ & 1017 & 1168 & 1163 \\
    $\tilde{u}_R$ & 946 & 1086 & 1089 \\
    $\tilde{d}_L$ & 1022 & 1173 & 1169 \\
    $\tilde{d}_R$ & 905 & 1047 & 1063 \\    
    $\tilde{b}_1$ & 894 & 1036  & 1053 \\
    $\tilde{b}_2$ & 929 & 1073 & 1085 \\
    $\tilde{t}_1$ & 704 & 831 & 879 \\
    $\tilde{t}_2$ & 957 & 1097 & 1107 \\
    $\tilde{\nu}_l$ & 564 & 621 & 556 \\
    $\tilde{\nu}_{\tau}$ & 562 & 619 & 555 \\
    $\tilde{e}_L$ & 569 & 626 & 561 \\
    $\tilde{e}_R$ & 478 & 497 & 403 \\
    $\tilde{\tau}_1$ & 473 & 492 & 399 \\
    $\tilde{\tau}_2$ & 568 & 625 & 561 \\ \hline
    $\tilde{\chi}_1^0$ & 395 & 464 & 459 \\
    $\tilde{\chi}_2^0$ & 439 & 514 & 508 \\
    $\tilde{\chi}_3^0$ & 530 & 595 & 562 \\
    $\tilde{\chi}_4^0$ & 571 & 640 & 621 \\
    $\tilde{\chi}_1^{\pm}$ & 433 & 506 & 496 \\
    $\tilde{\chi}_2^{\pm}$ & 568 & 636 & 618 \\ \hline
    $h^0$ & 114 & 115 & 114 \\
    $H^0$ & 766 & 852 & 783 \\
    $A^0$ & 765 & 851 & 783 \\
    $H^{\pm}$ & 770 & 856 & 787 \\ \hline \bottomrule \\\\\\
   \end{tabular}
   &
   \begin{tabular}{lrrr} \toprule\hline
    & Case 1 & Case 2 & Case 3 \\ \hline
    $\tilde{u}_L \to \tilde{g}\, u$ & $.23$ & $.23$ & $.20$ \\
    $\phantom{\tilde{u}_L} \to \tilde{\chi}_1^+\, d$
    & $.41$ & $.39$ & $.34$ \\
    $\phantom{\tilde{u}_L} \to \tilde{\chi}_2^0\, u$
    & $.21$ & $.21$ & $.18$ \\ \hline
    $\tilde{d}_L \to \tilde{g}\, d$ & $.25$ & $.24$ & $.22$\\
    $\phantom{\tilde{d}_L} \to \tilde{\chi}_1^-\, u$
    & $.41$ & $.35$ & $.29$ \\
    $\phantom{\tilde{d}_L} \to \tilde{\chi}_2^0\, d$
    & $.21$ & $.14$ & $.10$ \\ \hline
    $\tilde{u}_R \to \tilde{g}\, u$ & $.18$ & $.16$ & $.14$ \\
    $\phantom{\tilde{u}_R} \to \tilde{\chi}_1^0\, u$
    & $.67$ & $.71$ & $.63$ \\
    $\phantom{\tilde{u}_R} \to \tilde{\chi}_2^0\, u$
    & $.13$ & $.12$ & $.22$ \\ \hline
    $\tilde{d}_R \to \tilde{\chi}_1^0\, d$
    & $.83$ & $.84$ & $.69$ \\
    $\phantom{\tilde{d}_R} \to \tilde{\chi}_2^0\, d$
    & $.15$ & $.14$ & $.24$ \\ \hline
    $\tilde{g} \to \tilde{t}_1\,\bar{t} + \tilde{t}_1^*\,t$
    & $.98$ & $.98$ & \\
    $\phantom{\tilde{g}} \to \tilde{\chi}_2^+\,b\,\bar{t}
    + \tilde{\chi}_2^-\,t\,\bar{b}$
    & & & $.32$ \\
        $\phantom{\tilde{g}} \to \tilde{\chi}_{3,4}^0\,t\,\bar{t}$
    & & & $.32$ \\ \hline \hline
    $\tilde{\chi}_2^0 \to \tilde{\chi}_1^0\, q\, \bar{q}$ & $.42$ \\
    $\phantom{\tilde{\chi}_2^0} \to \tilde{\chi}_1^0\, b\, \bar{b}$ & $.12$ \\ 
    $\phantom{\tilde{\chi}_2^0} \to \tilde{\chi}_1^0\, l^+\, l^-$ &
	$.15$ \\
    $\phantom{\tilde{\chi}_2^0} \to \tilde{\chi}_1^0\, \tau^+\, \tau^-$ &
	$.10$ \\     \hline \hline
    $\tilde{\chi}_2^0 \to \tilde{l}^{\pm}_R\, l^{\mp}$ & & $.54$ \\
    $\phantom{\tilde{\chi}_2^0} \to \tilde{\tau}_1^{\pm}\, \tau^{\mp}$
    & & $.46$ \\ \hline
    $\tilde{l}^{\pm}_R \to \tilde{\chi}_1^0\, l^{\pm}$
    & & $1.00$ \\ \hline
    $\tilde{\tau}_1^{\pm} \to \tilde{\chi}_1^0\, \tau^{\pm}$
    & & $1.00$ \\ \hline \hline
    $\tilde{\chi}_2^0 \to \tilde{l}^{\pm}_R\, l^{\mp}$ & & & $.65$ \\
    $\phantom{\tilde{\chi}_2^0} \to \tilde{\tau}_1^{\pm}\, \tau^{\mp}$
    & & & $.35$ \\ \hline
    $\tilde{\chi}_1^0 \to \tilde{l}^{\pm}_R\, l^{\mp}$ & & & $.64$ \\
    $\phantom{\tilde{\chi}_1^0} \to \tilde{\tau}_1^{\pm}\, \tau^{\mp}$
    & & & $.36$ \\ \hline
    $\tilde{l}^{\pm}_R \to \tilde{\tau}_1\, l^{\pm}\, \tau$
    & & & $1.00$ \\ \hline \bottomrule
   \end{tabular}
  \end{tabular}
 \caption{Mass spectra and branching ratios of sparticles
 for three model points corresponding to three regions of $X + \bar{X}$
 messenger scenario. (See Figure \ref{fig:regionx}.)}
  \label{tab:parmx}
 \end{center}
\end{table}

\begin{table}
 \begin{center}
  \begin{tabular}{ccccccc} \toprule \hline
  $\Lambda_{\text{mess}}$ & $m_{\text{mess}}$ & $\tan\beta$
  & $\mathrm{sgn}\,(\mu)$ & $n_1$ & $n_2$ & $n_3$ \\
  (TeV) & (GeV) \\ \hline
  $80$ & $1.0 \times 10^{14}$ & $10$ & $+$ & $1/5$ & $3$ & $2$
			  \\ \hline \bottomrule
  \end{tabular}
 \end{center}
 \begin{center}
  \begin{tabular}{cc}
   \begin{tabular}{crcr} \toprule \hline
    $\quad \tilde{g} \quad$ & 1181 &
    $\quad \tilde{\chi}_1^0 \quad$ & 16 \\
    $\tilde{u}_L$ & 1326 & $\tilde{\chi}_2^0$ & 586 \\
    $\tilde{u}_R$ & 1165 & $\tilde{\chi}_3^0$ & 682 \\
    $\tilde{d}_L$ & 1331 & $\tilde{\chi}_4^0$ & 720 \\
    $\tilde{d}_R$ & 1163 & $\tilde{\chi}_1^{\pm}$ & 586 \\
    $\tilde{b}_1$ & 1150 & $\tilde{\chi}_2^{\pm}$ & 720 \\
    $\tilde{b}_2$ & 1221 & $h^0$ & 116 \\
    $\tilde{t}_1$ & 867 & $H^0$ & 954 \\
    $\tilde{t}_2$ & 1240 & $A^0$ & 954 \\
    $\tilde{\nu}_l$ & 680 & $H^{\pm}$ & 958 \\
    $\tilde{\nu}_{\tau}$ & 679 & & \\
    $\tilde{e}_L$ & 684 & & \\
    $\tilde{e}_R$ & 118 & & \\
    $\tilde{\tau}_1$ & 96 & & \\
    $\tilde{\tau}_2$ & 682 & & \\ \hline \bottomrule
   \end{tabular}
   &
   \begin{tabular}{lr} \toprule\hline
    $\tilde{u}_L \to \tilde{g}\, u$ & $.25$ \\
    $\phantom{\tilde{u}_L} \to \tilde{\chi}_1^+\, d$ & $.39$ \\
    $\phantom{\tilde{u}_L} \to \tilde{\chi}_2^0\, u$ & $.19$ \\ \hline
    $\tilde{d}_L \to \tilde{g}\, d$ & $.26$ \\
    $\phantom{\tilde{d}_L} \to \tilde{\chi}_1^+\, u$ & $.35$ \\
    $\phantom{\tilde{d}_L} \to \tilde{\chi}_2^0\, d$ & $.18$ \\ \hline
    $\tilde{u}_R \to \tilde{\chi}_1^0\, u$ & $1.00$ \\ \hline
    $\tilde{d}_R \to \tilde{\chi}_1^0\, d$ & $1.00$ \\ \hline
    $\tilde{g} \to \tilde{t}_1\,\bar{t} + \tilde{t}_1^*\,t$
    & $.97$ \\ \hline \hline
    $\tilde{\chi}_1^+ \to \tilde{\chi}_1^0\,W^+$ & $.91$ \\ \hline
    $\tilde{\chi}_2^0 \to \tilde{\chi}_1^0\,Z$ & $.30$ \\
    $\phantom{\tilde{\chi}_2^0} \to \tilde{\chi}_1^0\,h^0$ & $.54$ \\
    $\phantom{\tilde{\chi}_2^0} \to \tilde{l}_R^{\pm}\,l^{\mp}$ & $.04$
	\\ \hline
    $\tilde{l}_R^{\pm} \to \tilde{\chi}_1^0\,l^{\pm}$ & $1.00$
	\\ \hline \bottomrule
   \end{tabular}
  \end{tabular}
 \caption{Mass spectrum and branching ratios of sparticles
 for a model point of $Q + \bar{Q}$ messenger scenario.
 (See Figure \ref{fig:regionq}.)}
 \label{tab:parmq}  
 \end{center}
\end{table}

We pick three model points for $X + \bar{X}$ messenger scenario
corresponding to the three regions introduced above (Table \ref{tab:parmx})
and one model point for $Q + \bar{Q}$ messenger scenario
(Table \ref{tab:parmq}) to analyze the
LHC signals.
Table \ref{tab:parmx} and \ref{tab:parmq} show the resulting mass
spectra and branching ratios of sparticles in these model points.
In these points, the NLSP does not decay to the LSP gravitino inside
the detector.

As pointed out above, one of the most peculiar features of these scenarios
can be tested by measuring the masses of the bino and wino.
For the $X + \bar{X}$ messenger scenario, the mass splitting between
the bino and wino is very small.
On the other hand, for the $Q + \bar{Q}$ messenger scenario,
the mass of bino is much smaller than other sparticle masses.
So one of the most important tasks to distinguish these scenarios is
measuring the neutralino masses
\begin{equation}
 m_{\tilde{\chi}_1^0} \simeq M_1,\qquad
  m_{\tilde{\chi}_2^0} \simeq M_2.
\end{equation}
Of course we have to measure the mass of gluino
to confirm the relation among gaugino masses predicted by our scenarios.
But we do not argue the detailed reconstruction of the decay chain
for the gluino mass measurement in this paper. This is because the decay
mode of the gluino in the GMSB model is highly dependent on the parameters and
it can be very complicated.
As shown later, the gluino mass can be estimated if we assume that
the gluino mass is of the same order as the squark masses,
for example, by the $m_{T2}$ measurement and the largeness of the
cross section.

We figure out several features for three cases in
$X + \bar{X}$ messenger scenario and  one case in $Q + \bar{Q}$
messenger scenario.
\begin{enumerate}
\item $m_{\tilde W} : m_{\tilde g} \sim 1:2$ in both scenarios.
      As the result, the ratio $m_{\tilde W}/m_{\tilde q}$ becomes
      larger than in the usual scenario with the GUT
      relation. Roughly speaking, the hierarchy between colored sparticles
      and wino masses becomes milder.
\item $m_{\tilde B} \sim m_{\tilde W}$ in the $X+\bar X$ messenger scenario.
      In most of the interesting parameter region, the decay mode 
      $\chi_2^0\rightarrow Z\chi_1^0$ is closed, and therefore,
      the branching ratios of leptonic decay modes of $\chi_2^0$
      become comparatively large. The leptonic modes are important
      in obtaining meaningful information from the data in the LHC.
\item $m_{\tilde B} \ll m_{\tilde W}$ in the $Q+\bar Q$ messengner scenario.
\end{enumerate}
In the followings, we study how to catch these features from the LHC signals.

\subsection{$X + \bar{X}$ messenger scenario (Case 3: stau NLSP)}
\label{sec:x3}
In the region 3 of $X + \bar{X}$ messenger scenario, very peculiar signal
is expected because the NLSP becomes the right-handed stau.
The momentum and velocity of stau which goes out the detector can be
measured, and therefore, we can know the masses of various SUSY particles by
using the invariant mass technique and the above features can be tested.
We discuss the mass measurements for this case
in this subsection.

It has been studies how to catch the stau in the LHC in
\cite{Polesello:1999,Ambrosanio:2000ik,Ellis:2006vu,Rajaraman:2007ae}, 
and we carry out the smearing of the stau momentum and velocity to
reproduce the expected resolution in our simulation.
The resolutions for the momentum and
velocity are given as
\begin{equation}
 \sigma_{|\bm{p}|}(\mathrm{GeV}) =
  0.000118 \cdot |\bm{p}|^2 + 0.0002 \cdot
  \sqrt{m^2_{\tilde{\tau_1}} + |\bm{p}|^2} + 0.89
\end{equation}
and
\begin{equation}
 \sigma_{\beta} = 0.028 \cdot \beta^2.
\end{equation}
Then we can obtain the stau mass 
\begin{equation}
 m_{\tilde{\tau}_1} = \frac{|\bm{p}|}{\beta\gamma},\qquad
  \gamma \equiv (1 - \beta^2)^{-1/2}
\end{equation}
from the reconstructed momentum and velocity.
For the identification of stau, we use the cuts
\begin{itemize}
 \item $0.9 < \beta\gamma < 6.0$
 \item $|\eta| < 2.4$
 \item $p_T > 20\,\mathrm{GeV}$
\end{itemize}
in our analysis.
Here $\eta$ is the pseudo rapidity.

\begin{figure}
 \begin{center}
  \begin{tabular}{cc}
   \includegraphics[scale=0.75]{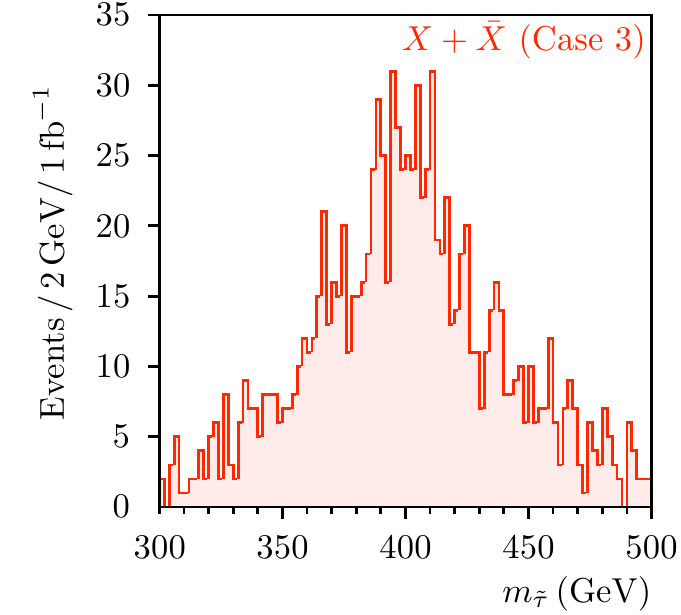} &
   \includegraphics[scale=0.75]{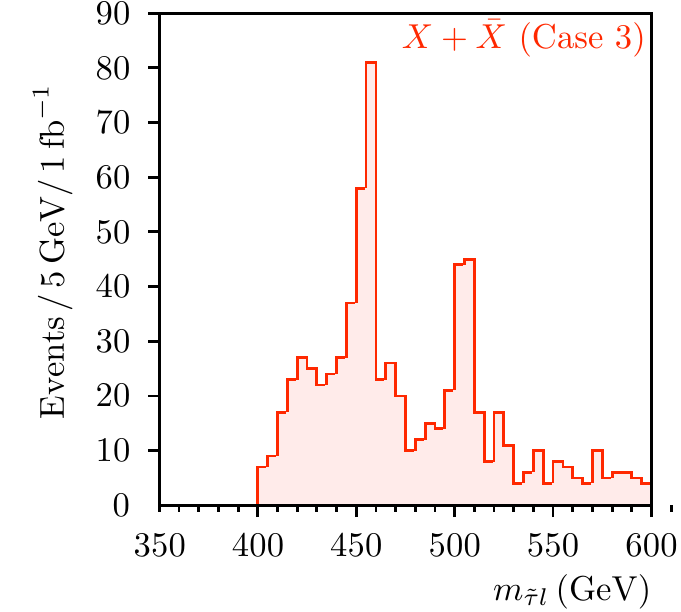}
   \includegraphics[scale=0.75]{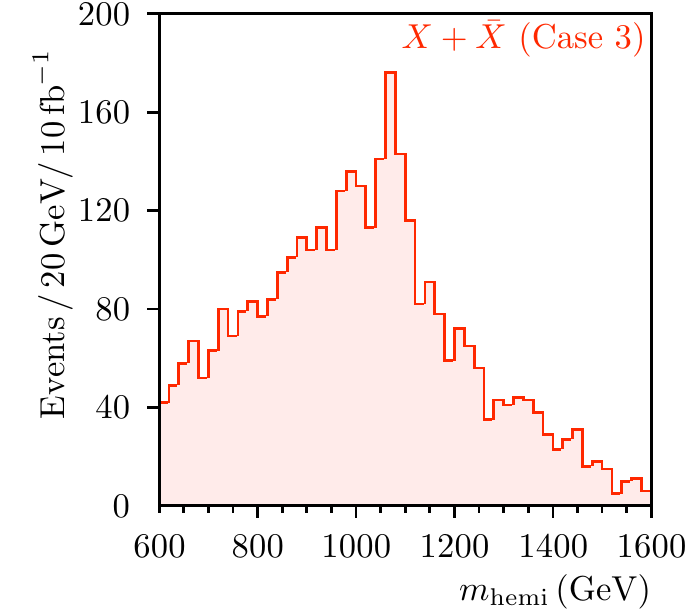}   
  \end{tabular}
 \caption{Left: NLSP stau invariant mass distribution for the case 3 of
  $X + \bar{X}$ messenger scenario.
  Center: $m_{\tilde{\tau}l}$ distribution.
  Right: Hemisphere invariant mass distribution.}
 \label{fig:x3_1}
 \label{fig:x3_2}
 \label{fig:x3_3}  
 \end{center}
\end{figure}

To measure the neutralino masses $m_{\tilde{\chi}_{1,2}^0}$,
we use the decay chain of
$\tilde{\chi}_{1,2}^0 \to \tilde{l}_R\,l_n \to
\tilde{\tau}_1\,l_n\,l_f\,\tau$.
Since mass difference between the lighter stau $\tilde{\tau}_1$ and
the first two generation of right-handed sleptons $\tilde{l}_R$ are
small in most of the cases we are interested in,
we use the following approximation:
\begin{equation}
 p_{\tilde{l}_R} = p_{\tilde{\tau}} + p_{l_f} + p_{\tau}
  \simeq p_{\tilde{\tau}_1}.
\end{equation}
So we can measure the masses of the bino-like and wino-like
neutralinos directly by taking the invariant mass of stau and
$e$ or $\mu$ as illustrated in Figure \ref{fig:x3_2}.
Although we can use the decay chain
$\tilde{\chi}_{1,2}^0 \to \tilde{\tau}_1\,\tau$ for the mass measurement
of neutralinos, by this measurement the events with $e$ or $\mu$ are
preferable because the momentum of $e$ or $\mu$ is less smeared
than that of $\tau$.
And we can confirm that the measured neutralinos are not higgsino-like
one because of the large branching ratio for
$\tilde{\chi}_{1,2}^0 \to \tilde{l}_R\,l$.

In this scenario, it is expected that we can measure the masses of
any kind of sparticles produced in each event.
This is because the momentum
of the arbitrary sparticle can be reconstructed by the momenta of
the stau and the SM particles.
So we can measure the mass of gluino and check the mass relation
among all the gauginos in principle.
Let us consider this issue in the rest of this subsection.

Although a large number of gluino are expected to be produced by
the process $p\,p \to \tilde{g}\,\tilde{g},\,\tilde{g}\,\tilde{q}$ and the
subsequent decay of squark, it is not always easy to distinguish gluino
from squark in event-by-event level.
Therefore we adopt the inclusive measurement of the invariant mass
of produced sparticles.
For this purpose, we use the hemisphere method suggested in
\cite{Ball:2007zza,Matsumoto:2006ws,Nojiri:2008vq}.
In this method, we sort the clusters into two hemispheres in each
event according to the following algorithm.
\begin{enumerate}
 \item We pick all the jets with $p_T > 50\,\mathrm{GeV}$, $|\eta| < 2$,
       leptons with $p_T > 15\,\mathrm{GeV}$, $|\eta| < 3$
       and two staus in each event.
       We use them as the clusters which compose two hemispheres
       corresponding to the pair-produced sparticles.
 \item We define the initial hemisphere axes $p_{\text{hemi}}^{(i)}$
       $(i = 1,2)$ by the momentum of two clusters.
       $p_{\text{hemi}}^{(1)}$ is defined as the momentum of the
       highest $p_T$ cluster.
       And $p_{\text{hemi}}^{(2)}$ corresponds to the momentum of
       the cluster which has the largest value of  $p_T \times \Delta R$
       where $\Delta R \equiv \sqrt{\Delta\eta^2 + \Delta\phi^2}$.
       Here $\Delta\eta \equiv \eta^{(1)} - \eta^{(2)}$,
       $\Delta\phi \equiv \phi^{(1)} - \phi^{(2)}$
       and $\phi$ is the azimuthal angle of the cluster.
 \item The cluster with momentum $p$ is belonging to
       the hemisphere 1 if it satisfies
       \begin{equation}
	d(p,p_{\text{hemi}}^{(1)}) < d(p,p_{\text{hemi}}^{(2)})
       \end{equation}
       and vise versa.
       Here $d(p_j,p_{\text{hemi}}^{(i)})$ is the Lund distance measure
       between the clusters with momentum $p_j$ and $p_{\text{hemi}}^{(i)}$,
       and it is defined by
       \begin{equation}
	d(p_j,p_{\text{hemi}}^{(i)}) =
	 (E_{\text{hemi}}^{(i)} - |\bm{p}_{\text{hemi}}^{(i)}|
	 \cos\theta_{ij})\frac{E_{\text{hemi}}^{(i)}}
	 {(E_{\text{hemi}}^{(i)} + E_j)^2}
       \end{equation}
       where $\theta_{ij}$ is the angle between $\bm{p}_j$ and
       $\bm{p}_{\text{hemi}}^{(i)}$.
 \item We redefine the hemisphere axis $p_{\text{hemi}}^{(i)}$ as
       the sum of the momenta of the clusters which belong to the
       hemisphere $i$.
 \item We repeat the step 3 and 4 until the classification of
       hemisphere converges.
\end{enumerate}
After this algorithm, we can obtain the invariant mass distribution
of each hemisphere $m_{\text{hemi}}^2 = p_{\text{hemi}}^2$.
If the assignment of hemisphere agrees with the true hemisphere,
$m_{\text{hemi}}$ corresponds to the mass of the pair produced
sparticle.
Since the true hemisphere should contain exactly one stau,
we reject the event where two staus are contained in one of the
hemispheres.

Using this algorithm, we illustrate the distribution of
$m_{\text{hemi}}$ in Figure \ref{fig:x3_3}.
We can see from this figure that both gluino and squark have
masses around $1000\,\mathrm{GeV}$ and the mass relation
$m_{\tilde{W}} : m_{\tilde{g}} \sim 1 : 2$ can be checked.

\subsection{Low luminosity analysis}
\label{sec:low}
In order to catch the features in the other cases of $X + \bar{X}$
messenger scenario and of $Q + \bar{Q}$ messenger scenario,
we discuss the several analyses which can be done in the LHC with
comparatively low luminosity.

\begin{figure}
 \begin{center}
   \includegraphics[scale=1.0]{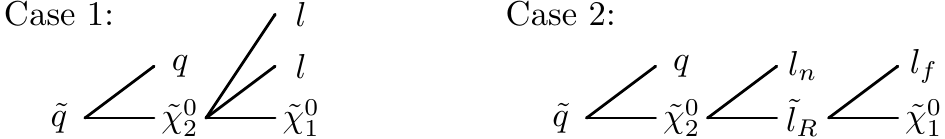}
 \caption{Decay chains which we use for the analysis in this paper.
 $\tilde{\chi}_2^0$ decays to $\tilde{\chi}_1^0$ via the off-shell
 slepton in the case 1. In the case 2, $\tilde{\chi}_2^0$ decays to on-shell
 slepton and we label the leptons which come from the decay of
 $\tilde{\chi}_2^0$ and $\tilde{l}_R$ as $l_n$ and $l_f$, respectively.}
 \label{fig:tree}    
 \end{center}
\end{figure}

Since we are interested in the measurement related to the masses of
the bino and wino, we make use of the characteristic decay modes
of these particles.
As noted in the comment of feature 2, 
for $X+\bar X$ messenger scenario,
the leptonic decay of wino-like neutralino $\tilde{\chi}_2^0$
to the lightest bino-like neutralino $\tilde{\chi}_1^0$
through on-shell or off-shell slepton $\tilde{l}_R$ is useful,
which are illustrated in Figure \ref{fig:tree}.

\begin{figure}
 \begin{center}
  \begin{tabular}{ccc}
   \includegraphics[scale=0.75]{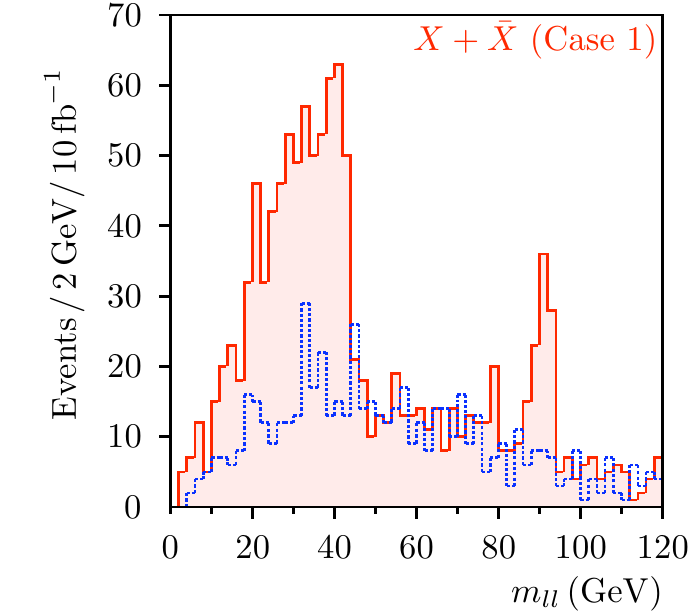} &
   \includegraphics[scale=0.75]{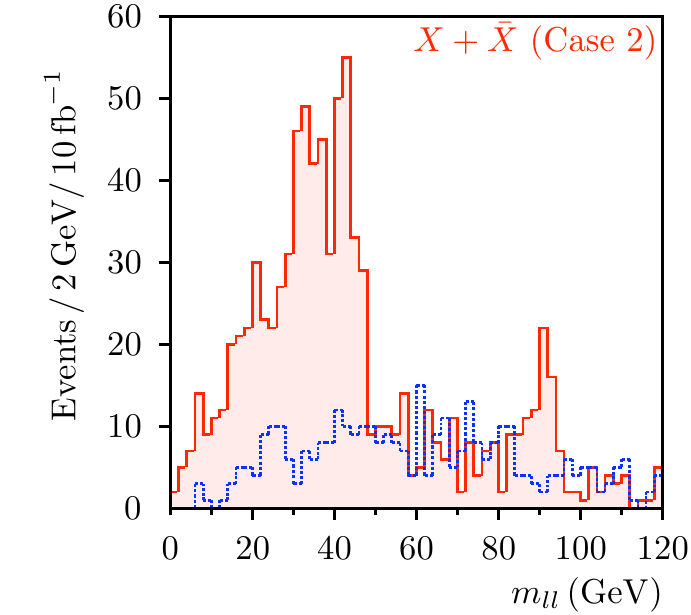} &
   \includegraphics[scale=0.75]{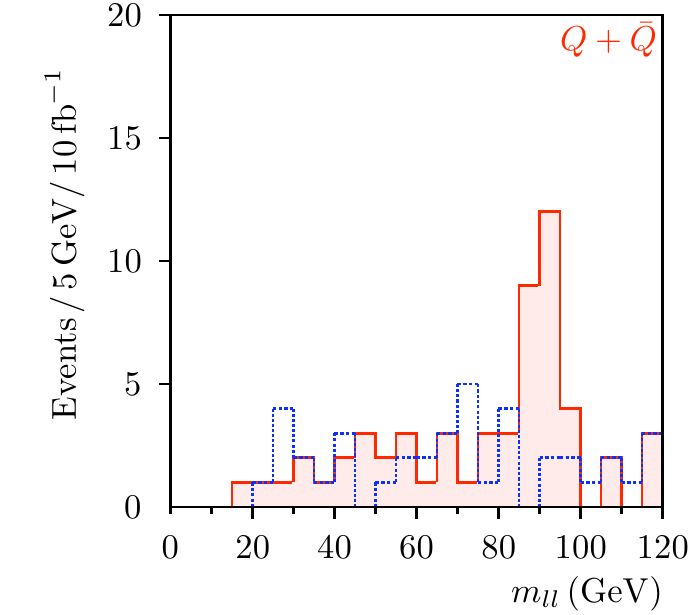}   
  \end{tabular}
 \caption{Dilepton invariant mass $m_{ll}$ distributions for
 $10\,\mathrm{fb}^{-1}$ of each model points.
  The bins illustrated by the dotted line represents the similar
  distribution using the opposite-sign and different flavor
  dileptons ($e^+\,\mu^- + \mu^+\,e^-$).}
 \label{fig:mll}     
 \end{center}
\end{figure}

First, we consider the invariant mass of a pair of leptons
coming from the decay of $\tilde{\chi}_2^0$.
As can be seen from Table \ref{tab:parmx}, in the $X+\bar X$ messenger
scenario, a large number of
$\tilde{\chi}_2^0$ is expected to be produced via the decay of
left-handed squark $\tilde{q}_L$.
We use a dilepton with opposite-sign and same-flavor
($e^+\,e^- + \mu^+\,\mu^-$) for making the invariant mass
and the distribution shown in Figure \ref{fig:mll} is obtained.
In this figure, the background is estimated by taking the similar
invariant mass of dilepton with opposite-sign and different-flavor
($e^+\,\mu^- + \mu^+\,e^-$).
Here, in order to reduce the SM background, we impose following event
cuts by using the transverse momenta $p_T$ \cite{:1999fq}
\begin{itemize}
 \item $p_{T}^{(1)} > 100\,\mathrm{GeV}$ and
       $p_{T}^{(2,3,4)} > 50\,\mathrm{GeV}$
 \item $M_{\text{eff}} \equiv p_T^{(1)} + p_T^{(2)} + p_T^{(3)} +
       p_T^{(4)} + E_T^{\text{miss}} > 400\,\mathrm{GeV}$
 \item $E_T^{\text{miss}} > \max\{100\,\mathrm{GeV},0.2M_{\text{eff}}\}$
 \item Two isolated leptons with $p_T^e > 20\,\mathrm{GeV}$ and
       $p_T^{\mu} > 5\,\mathrm{GeV}$       
\end{itemize}
where $p_T^{(i)}$ means the $i$-th largest $p_T$ of the jet in
each event and
$E_T^{\text{miss}} = \sqrt{(p_x^{\text{miss}})^2 + (p_y^{\text{miss}})^2}$.
Since the SM background is reduced successfully after
these cut, we generate only events of sparticle production for
our simulation \cite{:1999fq}.

We can see the rather small maximum value of invariant mass
 for both cases of the  $X + \bar{X}$ scenario
in Figure \ref{fig:mll}, which is caused by the feature 2, namely, 
$m_{\tilde B}\sim m_{\tilde W}$.
Actually, the maximum value of invariant mass allowed by
kinematics is given as
\begin{equation}
 \label{eq:x1ll}
 m_{ll}^{\max}[\text{Case 1}] = m_{\tilde{\chi}_2^0} - m_{\tilde{\chi}_1^0}
\end{equation}
in region 1, and 
\begin{equation}
 (m_{ll}^{\max}[\text{Case 2}])^2 = m_{\tilde{\chi}_2^0}^2
  \left(1 - \frac{m_{\tilde{l}_R}^2}{m_{\tilde{\chi}_2^0}^2}\right)
  \left(1 - \frac{m_{\tilde{\chi}_1^0}^2}{m_{\tilde{l}_R}^2}\right)  
\end{equation}
in region 2, which result in rather small maximum value of the invariant
mass calculated as
\begin{equation}
 m_{ll}^{\max}[\text{Case 1}] = 44\,\mathrm{GeV},\qquad
 m_{ll}^{\max}[\text{Case 2}] = 47\,\mathrm{GeV}.
\end{equation}
Unfortunately, the smallness of $m_{ll}^{\max}$ does not always mean the 
smallness of the mass splitting between the bino and wino.
For the case 2, if one of the relations
$m_{\tilde B}\sim m_{\tilde l}$ and $m_{\tilde l}\sim m_{\tilde W}$ is
satisfied, the maximum value of the invariant mass becomes small.
\begin{figure}
 \begin{center}
  \includegraphics[scale=0.75]{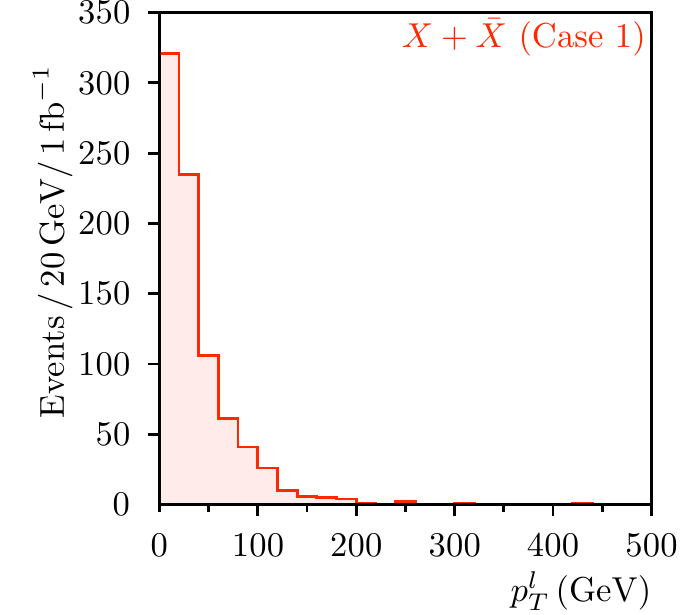}
  \includegraphics[scale=0.75]{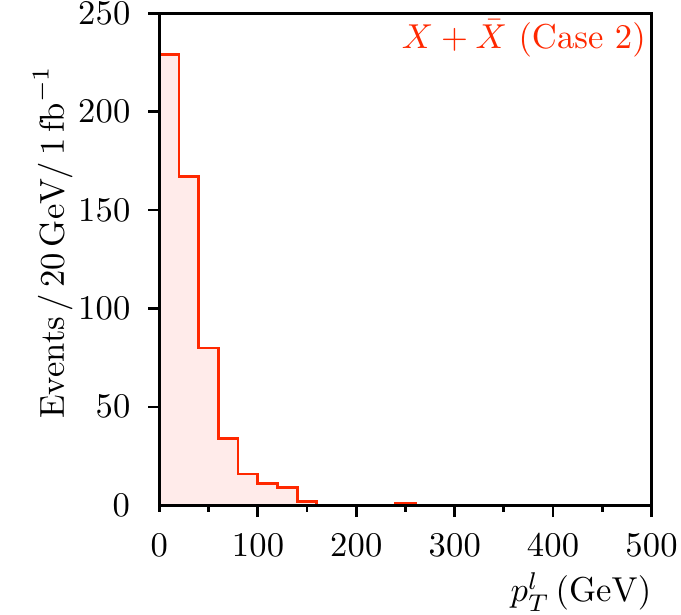}
  \includegraphics[scale=0.75]{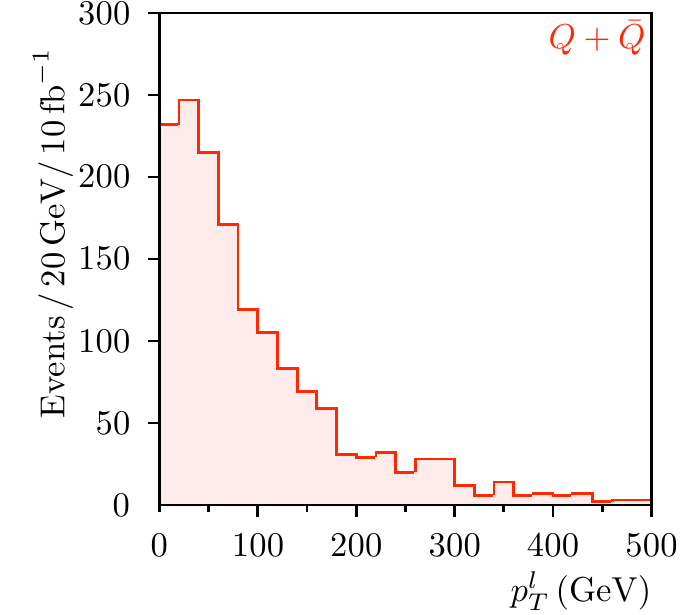}  
 \end{center}
 \caption{$p_T$ distributions of leptons emitted from sparticle decays.
 Each graph corresponds to $1\,\mathrm{fb}^{-1}$ for the case 1 of
 $X + \bar{X}$ scenario (left),
 $1\,\mathrm{fb}^{-1}$ for the case 2 of $X + \bar{X}$ scenario (center)
 and $10\,\mathrm{fb}^{-1}$ for $Q + \bar{Q}$ scenario (right).} 
 \label{fig:ptl} 
\end{figure}

To check that the mass difference between $\tilde{\chi}_1^0$ and
$\tilde{\chi}_2^0$ is small in $X + \bar{X}$ scenario,
we examine the $p_T$ distribution of leptons which come from the
decay shown in Figure \ref{fig:tree}.
Since the magnitude of $p_T$ of produced particles is strongly dependent 
on the mass difference among the sparticles, this can be a good 
signal to distinguish these scenarios. 
We can see from Figure \ref{fig:ptl} that these leptons
have relatively small $p_T$ in the $X+\bar X$ scenario and
large $p_T$ in the $Q+\bar Q$ scenario.
These are caused by $m_{\tilde B}\sim m_{\tilde W}$ in the $X+\bar X$ scenario
and $m_{\tilde B}\ll m_{\tilde W}$ in the $Q+\bar Q$ scenario.

Even in the models with the GUT relation for the gaugino masses,
such a small value of $m_{ll}^{\max}$ is possible if the gaugino
mass scale is small.
But such lighter gluino can be distinguished from heavier gluino
in $X + \bar{X}$ messenger scenario by measuring the cross section
and/or by the $m_{T2}$ method. As a reference model, we adopt the
minimal supergravity (mSUGRA) model with parameters
$m_{1/2} = 150\,\mathrm{GeV}$, $m_0 = 750\,\mathrm{GeV}$,
$A_0 = -100\,\mathrm{GeV}$, $\tan\beta = 10$ and
$\mathrm{sgn}\,(\mu) = +1$.
As shown later, though the distributions of $m_{ll}$ and
$p_T$ in the reference model are similar to those of $X + \bar{X}$
messenger scenario in Figure \ref{fig:mll} and \ref{fig:ptl},
the distribution of $m_{T2}$ and the cross section become
much different from those of $X + \bar{X}$ messenger scenario.
Let us remind the $m_{T2}$ method \cite{Lester:1999tx,Barr:2003rg,Barr:2009wu}.
When we consider the production process of sparticle pair which decay
into a pair of the NLSPs and a pair of the SM particles, 
we can make use of the $m_{T2}$ variable defined by
\begin{equation}
 \label{eq:mt2}
 m_{T2}(M_{\text{test}}) \equiv \min_{\bm{p}_T^{\text{miss}}
 = \sum_i\bm{p}_T^{\chi(i)}}
 \left[\max
  \left\{m_T(\overrightarrow{p}_T^{\text{vis}(1)},
   \overrightarrow{p}_T^{\chi(1)}),
   m_T(\overrightarrow{p}_T^{\text{vis}(2)},
   \overrightarrow{p}_T^{\chi(2)})\right\}\right].
\end{equation}
Here $\overrightarrow{p}_T$ is a $(2 + 1)$-dimensional
vector, $\overrightarrow{p}_T \equiv (E_T,\bm{p}_T)$, and 
$p_T^{\text{vis}(i)}$ is the transverse momentum
for the emitted observed particles.
$m_T$ is given by
\begin{equation}
 m_T^2(\overrightarrow{p}_T^{\text{vis}},\overrightarrow{p}_T^{\chi})
  \equiv (\overrightarrow{p}_T^{\text{vis}} + \overrightarrow{p}_T^{\chi})^2
  = m_{\text{vis}}^2 + M_{\text{test}}^2
  + 2(E_T^{\text{vis}}\cdot E_T^{\chi}
  - \bm{p}_T^{\text{vis}} \cdot \bm{p}_T^{\chi})
\end{equation}
where $E_T^{\text{vis}} \equiv
\sqrt{m_{\text{vis}}^2 + |\bm{p}^{\text{vis}}_T|^2}$
and $E_T^{\chi} \equiv \sqrt{M_{\text{test}}^2 + |\bm{p}^{\chi}_T|^2}$.

Now we consider the process
$p\,p \to \tilde{q}\,\tilde{q} \to
\tilde{\chi}_1^0\,q\,\tilde{\chi}_1^0\,q$.
In this process, $m_{\text{vis}}^{(1)} \simeq m_{\text{vis}}^{(2)}
\simeq 0$ and the maximum value of $m_{T2}$ is given by
\begin{equation}
 \label{eq:x1mt2}
 m_{T2}^{\max}(M_{\text{test}}) =
  \frac{m_{\tilde{q}}^2 - m_{\tilde{\chi}_1^0}^2}{2m_{\tilde{q}}}
  + \sqrt{
  \left(
   \frac{m_{\tilde{q}}^2 - m_{\tilde{\chi}_1^0}^2}{2m_{\tilde{q}}}
  \right)^2 + M_{\text{test}}^2}
\end{equation}
as a function of the test mass $M_{\text{test}}$.
Therefore, we can obtain the rough value of the colored sparticle masses
by this analysis.
\begin{figure}
 \begin{center}
  \begin{tabular}{ccc}
   \includegraphics[scale=0.75]{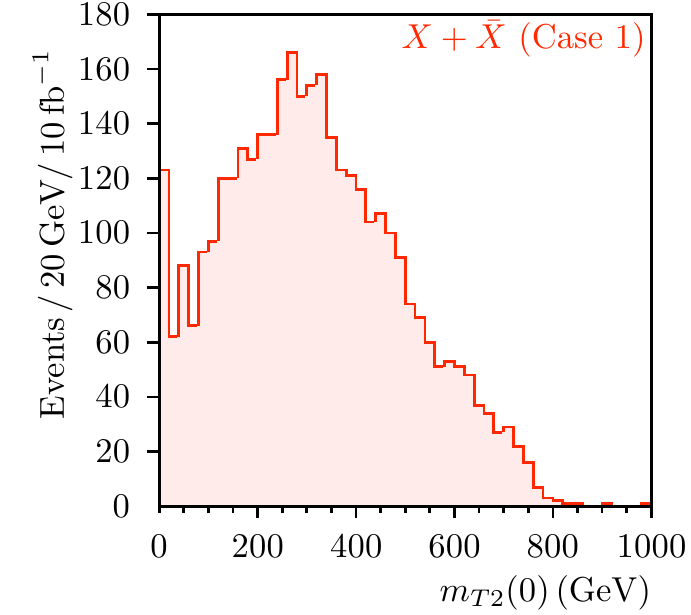} &
   \includegraphics[scale=0.75]{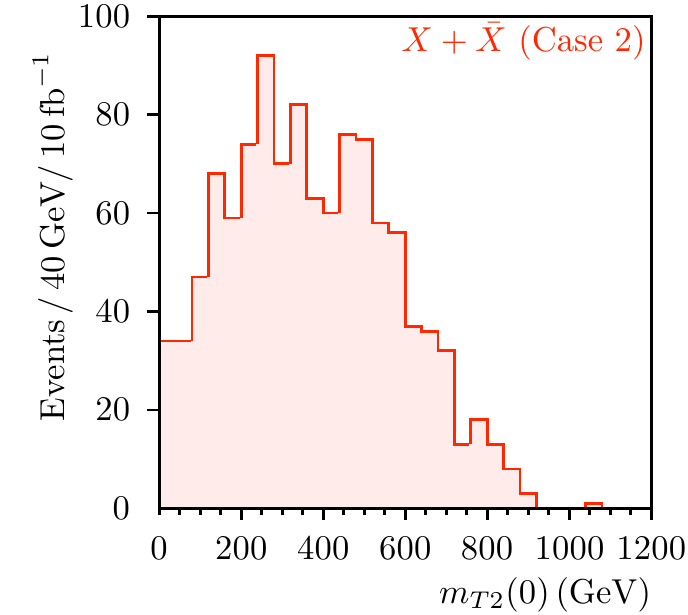} &
   \includegraphics[scale=0.75]{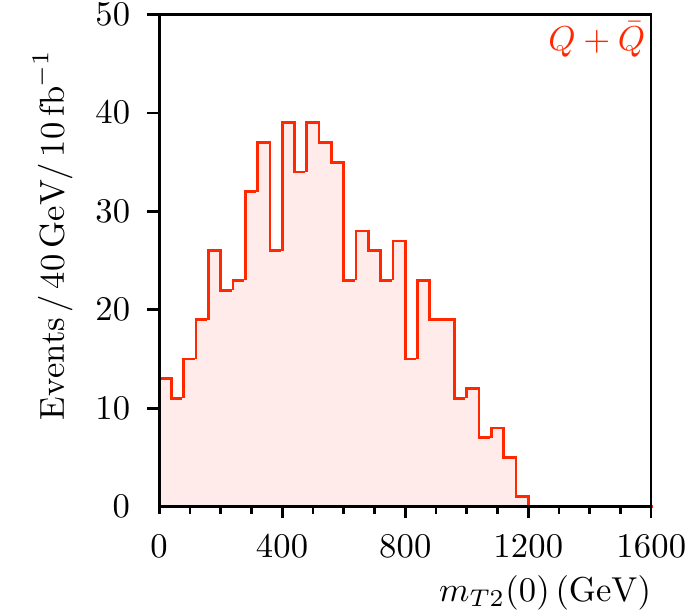}
  \end{tabular}
 \caption{$m_{T2}$ distributions for $M_{\text{test}} = 0$.}
 \label{fig:mt2}    
 \end{center}
\end{figure}

For the analysis of $m_{T2}$, we use the event cuts
\begin{itemize}
 \item Two jets with $p_T > 100\,\mathrm{GeV}$
 \item $M_{\text{eff}} \equiv p_T^{(1)} + p_T^{(2)} + E_T^{\text{miss}}
       > 400\,\mathrm{GeV}$
 \item $E_T^{\text{miss}} > \max\{100\,\mathrm{GeV},0.2M_{\text{eff}}\}$
 \item No lepton
\end{itemize}
instead of the usual cut for the SM events introduced above.
Since \eqref{eq:x1mt2} is satisfied for any fixed value of
$M_{\text{test}}$, here we set $M_{\text{test}} = 0$ for
our analysis. Then \eqref{eq:x1mt2} becomes the following simple form:
\begin{equation}
 \label{eq:mt20}
 m_{T2}^{\max}(0) =
  \frac{m_{\tilde{q}}^2 - m_{\tilde{\chi}_1^0}^2}{m_{\tilde{q}}}
\end{equation}
for the process of squark pair production.
The distribution of this quantity is shown in Figure \ref{fig:mt2}.
If the mass of the LSP is very small compared with the mass of
squark, we can interpret $m_{T2}^{\max}(0)$ as the mass scale of
squark.
In fact, this is the case for $Q + \bar{Q}$ messenger scenario.
However, the mass hierarchy of sparticles is small in $X + \bar{X}$
messenger scenario and the effect of $m_{\tilde{\chi}_1^0}$ is
non-negligible.
The theoretical values of $m_{T2}^{\max}(0)$ are
\begin{equation}
 \label{eq:x1chi}
 m_{T2}^{\max}(0)[\text{Case 1}] = 844\,\mathrm{GeV},\qquad
 m_{T2}^{\max}(0)[\text{Case 2}] = 904\,\mathrm{GeV},
\end{equation}
here we approximate $m_{\tilde q}=1000$ GeV for the case 1 and
$m_{\tilde q}=1100$ GeV for the case 2.
By this analysis, we can obtain the evidence of milder hierarchy 
between the masses of $\chi_2^0$ and colored sparticle
if we know $m_{\tilde{\chi}_2^0} \simeq 500\,\mathrm{GeV}$.
Unfortunately, we do not find the scale of $m_{\tilde{\chi}_2^0}$
by the analysis in this subsection and it needs further detailed analysis.

\begin{figure}
 \begin{center}
  \begin{tabular}{ccc}
   \includegraphics[scale=0.75]{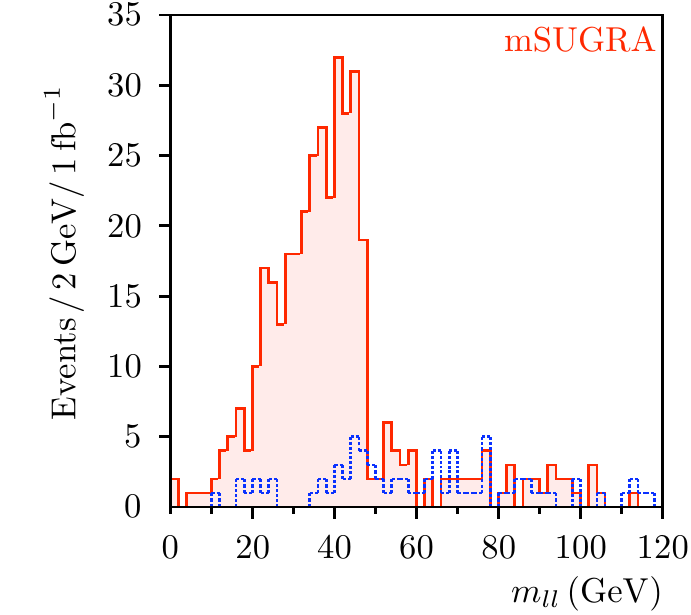} &
   \includegraphics[scale=0.75]{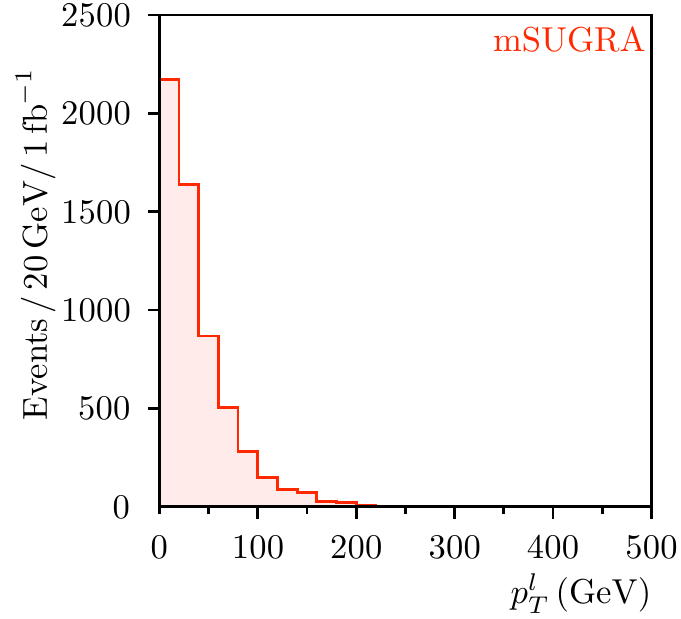} &
   \includegraphics[scale=0.75]{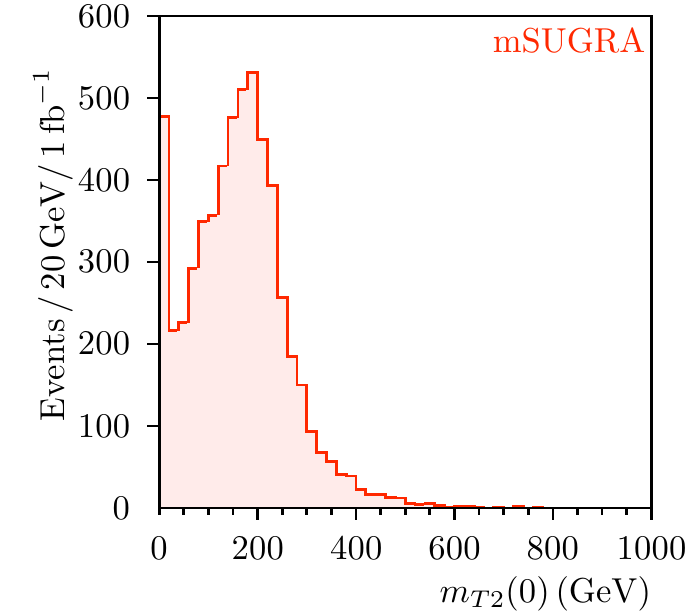}
  \end{tabular}
 \caption{The measurements for the mSUGRA model point, where
  $m_{1/2} = 150\,\mathrm{GeV}$, $m_0 = 750\,\mathrm{GeV}$,
  $A_0 = -100\,\mathrm{GeV}$, $\tan\beta = 10$ and
  $\mathrm{sgn}\,(\mu) = +1$.
  Left: Dilepton invariant mass $m_{ll}$ distribution.
  The bins illustrated by the dotted line represents the similar
  distribution using the opposite-sign and different flavor dileptons
  ($e^+\,\mu^- + \mu^+\,e^-$).
  Center: $p_T$ distribution of leptons emitted from sparticle decays.
  Right: $m_{T2}$ distribution for $M_{\text{test}} = 0$.}
 \label{fig:sugra}    
 \end{center}
\end{figure}

In the reference model, the distributions $m_{ll}$, $p_T$ and $m_{T2}$
are represented in Figure \ref{fig:sugra}.
The distribution of $m_{T2}$ is much different from those in
Figure \ref{fig:mt2}, although the distribution of $m_{ll}$ and
$p_T$ is similar to those of $X + \bar{X}$ messenger scenario
in Figure \ref{fig:mll} and \ref{fig:ptl}.
And the cross section becomes much larger than in $X + \bar{X}$ scenario.
Since the distribution of $m_{T2}$ and the cross section in
$X + \bar{X}$ messenger scenario show the much larger mass scale
of the colored particle than the gluino mass obtained by the GUT
relation, it is suggested that the GUT relation is not satisfied.

\subsection{$X + \bar{X}$ messenger scenario (Case 1, 2: neutralino
   NLSP)}
\label{sec:x12}   
In the cases 1 and 2 of $X + \bar{X}$ messenger scenario,
the neutralino is the
NLSP and we focus on the decay chain shown in Figure \ref{fig:tree}.
Although these two cases give similar signals, the decay modes of
the wino-like neutralino $\tilde{\chi}_2^0$ are different.
In the case 1, $\tilde{\chi}_2^0$ undergoes three-body decay
through off-shell slepton, while 
$\tilde{\chi}_2^0$ decays to the right-handed slepton
$\tilde{l}_R$, which decays to $\tilde{\chi}_1^0$ subsequently,
in the case 2.
As seen in the previous section, 
the small values of $m_{ll}^{\rm max}$ and $p_T^l$ indicate that
$m_{\tilde \chi_2^0}-m_{\tilde\chi_1^0}$ is small, but this may not
mean that $m_{\tilde \chi_2^0}\sim m_{\tilde\chi_1^0}$ because
the possibility may be still alive 
that the absolute value of the neutralino mass scale is small.
In order to reject the possibility, we try to show the relation
$m_{\tilde \chi_2^0}\sim m_{\tilde\chi_1^0}$ by measuring the invariant 
mass $m_{jl(u)}$ of a jet emitted from the squark $\tilde q$ and one of 
two leptons in the decay of $\tilde{\chi}_2^0$,
although the large luminosity is required 
for this analysis. Since there are two leptons in each event, we include
two invariant masses $m_{jl(u)}$ for each event in the distribution. 
In the case 1, the maximum value of $m_{jl(u)}$ is obtained as
\begin{equation}
 \label{eq:mjlu1}
 (m_{jl(u)}^{\max}[\text{Case 1}])^2 = m_{\tilde{q}}^2
    \left(1 - \frac{m_{\tilde{\chi}_2^0}^2}{m_{\tilde{q}}^2}\right)
    \left(1 - \frac{m_{\tilde{\chi}_1^0}^2}{m_{\tilde{\chi}_2^0}^2}\right),
\end{equation}
which is predicted to be 392 GeV for $m_{\tilde{q}} = 1000\,\mathrm{GeV}$.  
Note that this predicted value is much smaller
than $m_{T2}^{\rm max}(0)\sim 844$ GeV. This indicates that   
$m_{\tilde \chi_2^0}\sim m_{\tilde\chi_1^0}$ unless 
$m_{\tilde q}\sim m_{\tilde \chi_2^0}$.
Note that in the case 2, there are two kinds
of leptons in the decay because there is an on-shell slepton $\tilde l_R$
produced by the decay of $\tilde\chi_2^0$. 
Here we label two leptons emitted from 
$\tilde{\chi}_2^0$ and $\tilde{l}_R$
as ``near''-lepton $l_n$ and ``far''-lepton $l_f$, respectively,
as shown in Figure \ref{fig:tree}.
Since we cannot distinguish
$l_n$ with $l_f$ in event-by-event level,
we consider the quantity
\begin{equation}
 \label{eq:mjlu}
 m_{jl(u)} \equiv m_{jl_n} \cup m_{jl_f}
\end{equation}
suggested by \cite{Burns:2009zi,Matchev:2009iw}.
$m_{jl_n}$($m_{jl_f}$) means the invariant mass of a jet emitted from
squark $\tilde{q}$ and a lepton $l_n$($l_f$).
Then $m_{jl(u)}$ gives a combined distribution of $m_{jl_n}$ and
$m_{jl_f}$.
The important point is that the analysis is completely the same as
in the case 1,
and we do not have to distinguish $l_n$ and $l_f$ at event-by-event
level. The maximum value of the distribution of $m_{jl(u)}$ becomes
\begin{equation}
 \max\{m_{jl_n}^{\max},m_{jl_f}^{\max}\},
\end{equation}
where
\begin{equation}
 \label{eq:mjln}
 (m_{jl_n}^{\max})^2 = m_{\tilde{q}}^2
  \left(1 - \frac{m_{\tilde{\chi}_2^0}^2}{m_{\tilde{q}}^2}\right)
  \left(1 - \frac{m_{\tilde{l}_R}^2}{m_{\tilde{\chi}_2^0}^2}\right)
\end{equation}
and
\begin{equation}
 \label{eq:mjlf}
 (m_{jl_f}^{\max})^2 = m_{\tilde{q}}^2
  \left(1 - \frac{m_{\tilde{\chi}_2^0}^2}{m_{\tilde{q}}^2}\right)
  \left(1 - \frac{m_{\tilde{\chi}_1^0}^2}{m_{\tilde{l}_R}^2}\right).
\end{equation}
In the case 2, $m_{jl(u)}^{\max}$ is predicted to be 369 GeV because
$m_{jl_n}^{\max}=262$ GeV and $m_{jl_f}^{\max}=369$ GeV 
for $m_{\tilde{q}} = 1150\,\mathrm{GeV}$. Again, this 
predicted value is much smaller than $m_{T2}^{\rm max}(0)\sim 904$ GeV,
and it indicates that  $m_{\tilde \chi_2^0}\sim m_{\tilde\chi_1^0}$ 
because $m_{jl_n}^{\max}\ll m_{T2}^{\rm max}(0)$ means 
$m_{\tilde \chi_2^0}\sim m_{\tilde l_R}$ and 
$m_{jl_f}^{\max}\ll m_{T2}^{\rm max}(0)$ means 
$m_{\tilde \chi_1^0}\sim m_{\tilde l_R}$ 
unless 
$m_{\tilde q}\sim m_{\tilde \chi_2^0}$.
Therefore, if $m_{jl(u)}^{\rm max}$ is much smaller than $m_{T2}^{\rm max}(0)$,
$m_{\tilde \chi_2^0}\sim m_{\tilde\chi_1^0}$ can be shown.

\begin{figure}
 \begin{center}
  \begin{tabular}{ccc}
   \includegraphics[scale=1.0]{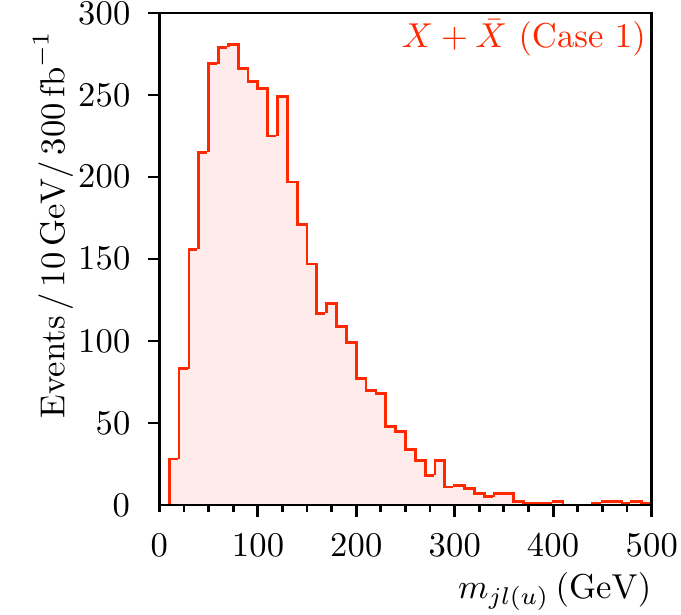} &
   \includegraphics[scale=1.0]{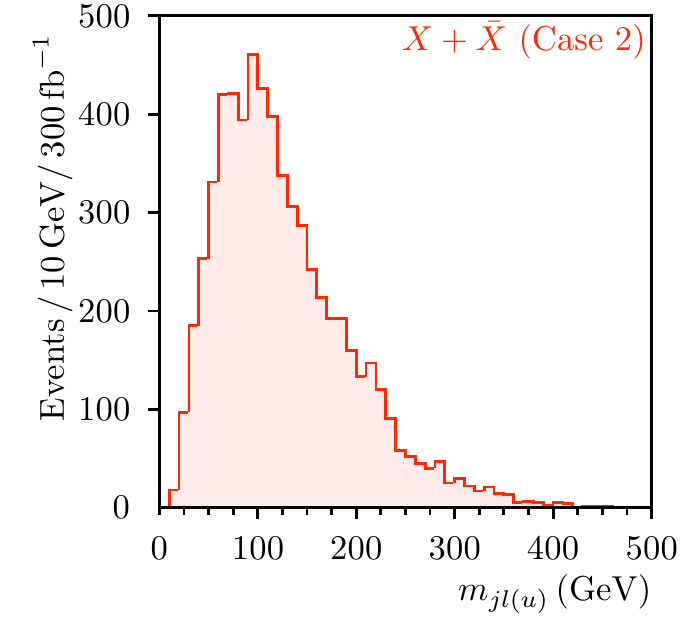}
  \end{tabular}
  \caption{$m_{jl(u)}$ distributions for $300\,\mathrm{fb}^{-1}$ of
  the case 1 (left) and case 2 (right) of $X + \bar{X}$ scenario.}
  \label{fig:x2_1}
  \label{fig:x2_2}    
 \end{center}
\end{figure}

In our simulation, we impose the cuts for the standard model background
as in the section \ref{sec:low} and use the dilepton whose invariant mass
$m_{ll}$ is less than $50\,\mathrm{GeV}$ and a jet with $p_T$
larger than $100\,\mathrm{GeV}$.
There are, however, many background of jets coming from other decays
of colored sparticles, such as
$\tilde{g} \to \tilde{t}_1\,\bar{t} \to \tilde{\chi}_2^+\,b\,\bar{t}$.
To reduce these background, we impose another event cut that there is no
$b$-tagged jet in each event. Here we assume $60\%$ tagging efficiency
of $b$-jet.
Then we make two invariant masses $m_{jl_i}$ ($i = 1, 2$) for all the
possible jets in each event.
And we choose a jet which minimizes $\max\{m_{jl_1},m_{jl_2}\}$ among
these jets.
In this way we can obtain the distribution of $m_{jl(u)}$
which consists of the combined distribution of $m_{jl_1}$ and
$m_{jl_2}$.
The above predicted values are roughly consistent with the measured 
values obtained from Figure \ref{fig:x2_1}.

Note that in the case 1, 
taking account of three relations 
\eqref{eq:x1ll}, \eqref{eq:mt20} and \eqref{eq:mjlu1}
together, we can obtain the masses of squark $\tilde{q}$ and neutralinos
$\tilde{\chi}_{1,2}^0$, in principle.
We will return to this point in the end of this subsection.

In order to determine the mass spectrum of sparticles in the case 2, 
more measurements are needed. As suggested by \cite{Matchev:2009iw},
$\min\{m_{jl_n}^{\max},m_{jl_f}^{\max}\}$ can be obtained
from the distribution of $m_{jl(u)}$
by observing the structure with the intermediate endpoints in the case 2.
Actually the predicted value roughly agrees with the measured value
in Figure \ref{fig:x2_2}.

\begin{figure}
 \begin{center}
  \includegraphics[scale=1.0]{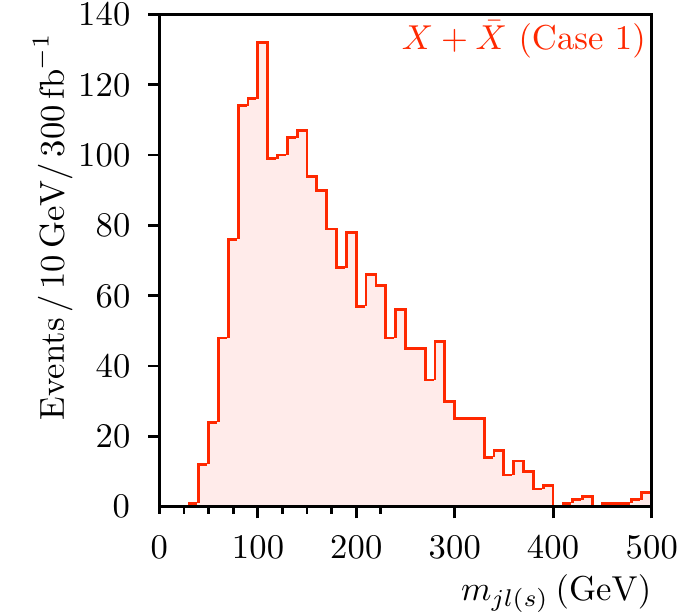}
  \includegraphics[scale=1.0]{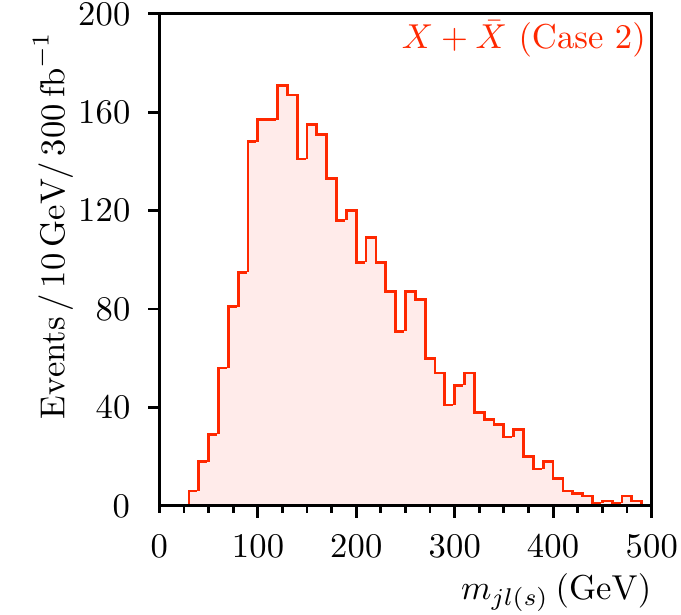}
  \caption{$m_{jl(s)}$ distributions for $300\,\mathrm{fb}^{-1}$ of
  the case 1 (left) and case 2 (right) of $X + \bar{X}$ scenario.}
  \label{fig:x2_3}    
 \end{center}
\end{figure}

Although we do not know which of the two measured endpoints of
$m_{jl(u)}$ in the case 2 corresponds to $m_{jl_n}$ ($m_{jl_f}$),
it is shown in \cite{Matchev:2009iw} that we can determine the
masses of neutralinos without ambiguity by introducing the quantity
\begin{equation}
 m_{jl(s)}^2 \equiv m_{jl_n}^2 + m_{jl_f}^2.
\end{equation}
The maximum value of this quantity is obtained as
\begin{align}
 (m_{jl(s)}^{\max}[\text{Case 2}])^2 &= (m_{jl_n}^{\max})^2
 + \frac{m_{\tilde{l}_R}^2}{m_{\tilde{\chi}_2^0}^2}(m_{jl_f}^{\max})^2
 \notag \\
 &= m_{\tilde{q}}^2
 \left(1 - \frac{m_{\tilde{\chi}_2^0}^2}{m_{\tilde{q}}^2}\right)
 \left(1 - \frac{m_{\tilde{\chi}_1^0}^2}{m_{\tilde{\chi}_2^0}^2}\right).
\end{align}
in the case 2.
Then we impose a similar event selection as in the previous one
and make an invariant mass by a dilepton with $m_{ll} < 50\,\mathrm{GeV}$
and a jet with $p_T > 100\,\mathrm{GeV}$.
Among all the possible choice of a jet, we pick the one which minimizes
$m_{jl(s)}$ in each events.
And the result of simulation is shown in
Figure \ref{fig:x2_3}, whereas the theoretical value is calculated as
\begin{equation}
 m_{jl(s)}^{\max}[\text{Case 2}] = 442\,\mathrm{GeV}.
\end{equation}
From these quantities, we can obtain the neutralino masses without
any ambiguity in the case 2 \cite{Matchev:2009iw}. By denoting that
\begin{equation}
 A \equiv m_{ll}^{\max},\quad
  B \equiv \max\{m_{jl_n}^{\max},m_{jl_f}^{\max}\},\quad
  C \equiv \min\{m_{jl_n}^{\max},m_{jl_f}^{\max}\},\quad
  D \equiv m_{jl(s)}^{\max},
\end{equation}
the masses of neutralinos are written as
\begin{equation}
 m_{\tilde{\chi}_1^0}[\text{Case 2}] =
  \frac{A\sqrt{(D^2 - B^2)(D^2 - C^2)}}{B^2 + C^2 - D^2},\qquad
  m_{\tilde{\chi}_2^0}[\text{Case 2}] = \frac{ABC}{B^2 + C^2 - D^2}.
\end{equation}
Moreover, the squark mass is also obtained as
\begin{equation}
m_{\tilde q}[\text{Case 2}] =
\frac{BC\sqrt{A^2+B^2+C^2-D^2}}{B^2+C^2-D^2}.
\end{equation}
If we take $A = 50\,\mathrm{GeV}$, $B = 380\,\mathrm{GeV}$,
$C = 240\,\mathrm{GeV}$ and $D = 440\,\mathrm{GeV}$,
we obtain $m_{\tilde{\chi}_1^0} \simeq 490\,\mathrm{GeV}$,
$m_{\tilde{\chi}_2^0} \simeq 540\,\mathrm{GeV}$
and $m_{\tilde{q}} \simeq 1130\,\mathrm{GeV}$,
which are in good agreement with the real
values in Table \ref{tab:parmq}.
The measured value of $m_{T2}^{\max}(0)$ can be used to check the
consistency.

We can also calculate the maximum value of $m_{jl(s)}$
for the case 1 as
\begin{align}
 \label{eq:mjll}
 m_{jl(s)}^{\max}[\text{Case 1}] &= m_{jll}^{\max}[\text{Case 1}]
  = m_{\tilde{q}}^2
 \left(1 - \frac{m_{\tilde{\chi}_2^0}^2}{m_{\tilde{q}}^2}\right)
 \left(1 - \frac{m_{\tilde{\chi}_1^0}^2}{m_{\tilde{\chi}_2^0}^2}\right)
 \notag \\
 &= m_{jl(u)}^{\max}[\text{Case 1}].
\end{align}
where $m_{jll}$ means the invariant mass by the jet and dilepton.
The first equality can be shown by the trivial relation
\begin{equation}
 m_{jl(s)}^2 = m_{jll}^2 - m_{ll}^2
\end{equation}
and the fact that $m_{jll}$ is maximized when $m_{ll} = 0$
as long as
$m_{\tilde{\chi}_2^0}/m_{\tilde{\chi}_1^0}
< m_{\tilde{q}}/m_{\tilde{\chi}_2^0}$ is satisfied \cite{Burns:2009zi}.
Let us determine the masses of squark and neutralinos
$\tilde{\chi}_{1,2}^0$.
For example, if we take $m_{ll}^{\max} = 45\,\mathrm{GeV}$,
$m_{T2}^{\max}(0) = 800\,\mathrm{GeV}$ and
$m_{jl(u)}^{\max} = m_{jl(s)}^{\max} = 400\,\mathrm{GeV}$,
then we can obtain $m_{\tilde{\chi}_1^0} \simeq 330\,\mathrm{GeV}$,
$m_{\tilde{\chi}_2^0} \simeq 370\,\mathrm{GeV}$ and
$m_{\tilde{q}} \simeq 920\,\mathrm{GeV}$
which are not far away from the real values in Table \ref{tab:parmx}.

\subsection{$Q + \bar{Q}$ messenger scenario}
\label{sec:q}
\begin{figure}
 \begin{center}
   \includegraphics[scale=1.0]{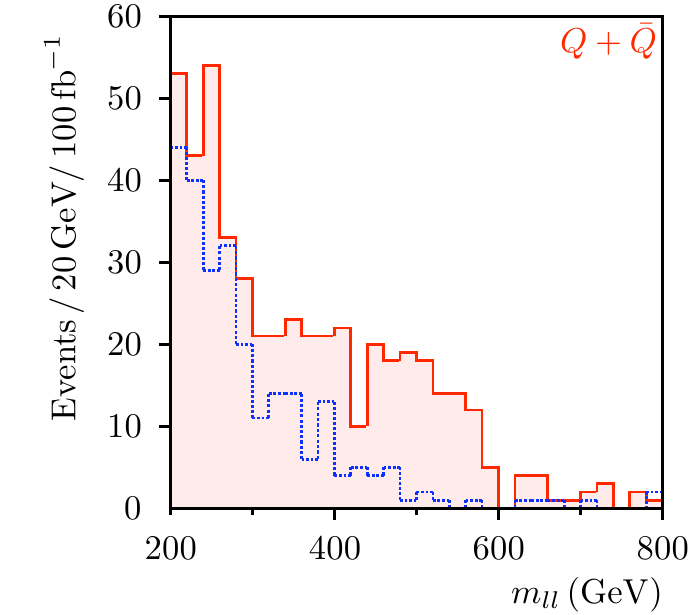}
   \includegraphics[scale=1.0]{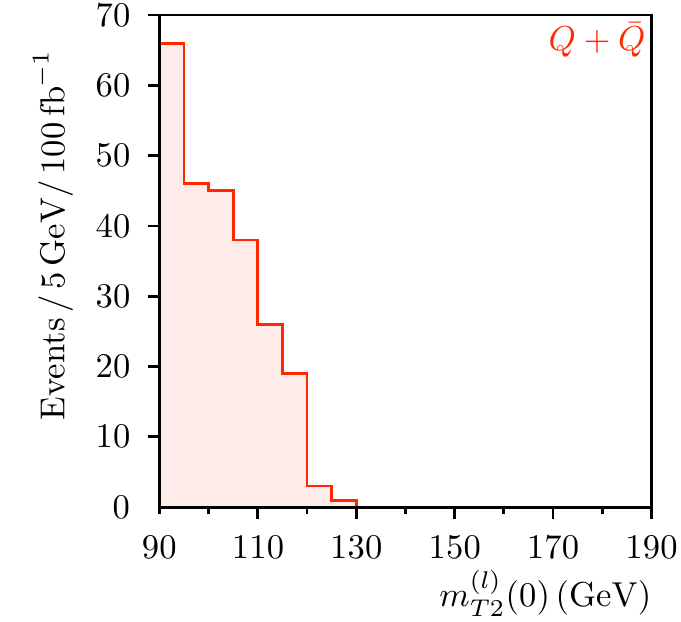}
 \caption{Left: Dilepton invariant mass $m_{ll}$ distribution for
  $100\,\mathrm{fb}^{-1}$ of $Q + \bar{Q}$ messenger scenario.
  The bins illustrated by the dotted line represents the similar
  distribution using the opposite-sign and different flavor
  dileptons ($e^+\,\mu^- + \mu^+\,e^-$).
  Right: leptonic $m_{T2}$ distribution for $100\,\mathrm{fb}^{-1}$
  of $Q + \bar{Q}$ messenger scenario.}
  \label{fig:q_2}  
 \end{center}
\end{figure}

In $Q + \bar{Q}$ messenger scenario, we can see the dilepton signal
if we collect a large number of events.
The maximum value of $m_{ll}$ is given as
\begin{equation}
 \label{eq:mllq}
 (m_{ll}^{\max})^2 = m_{\tilde{\chi}_2^0}^2
  \left(1 - \frac{m_{\tilde{l}_R}^2}{m_{\tilde{\chi}_2^0}^2}\right)
  \left(1 - \frac{m_{\tilde{\chi}_1^0}^2}{m_{\tilde{l}_R}^2}\right)  
\end{equation}
and the predicted value becomes
\begin{equation}
 m_{ll}^{\max} = 569\,\mathrm{GeV}
\end{equation}
in this model point.
The result of simulation corresponding to $100\,\mathrm{fb}^{-1}$
is shown in Figure \ref{fig:q_2} and
the measured value is consistent with the predicted value.
Since $m_{ll}^{\max}$ gives the lower bound of the mass of 
$\tilde{\chi}_2^0$, such a large value of $m_{ll}^{\max}$ indicates
the large $m_{\tilde{\chi}_2^0}$. In the section \ref{sec:low},
we have already the mass scale of the heaviest colored particle,
which is predicted as $1200\,\mathrm{GeV}$,
by $m_{T2}$ analysis for the colored sparticle pair production. 
These signals mean that the mass ratio 
$m_{\tilde{\chi}_2^0}/m_{\tilde{g}}$ is roughly larger than 1/2, and 
therefore, the hierarchy between the gluino and wino masses is milder
than that in the models with GUT relation. 

We can also use the $m_{T2}$ analysis for the right-handed slepton pair
production, because the right-handed slepton also has a smaller mass
compared with the other sparticles.
Using the two leptons emitted from a pair of sleptons, we obtain
the maximum value of $m_{T2}$ variable for
$M_{\text{test}} = 0\,\mathrm{GeV}$ as
\begin{equation}
 m_{T2}^{(l)\max}(0) =
  \frac{m_{\tilde{l}_R}^2 - m_{\tilde{\chi}_1^0}^2}{m_{\tilde{l}_R}}
  \label{mt2q}
\end{equation}
To select the process
$p\,p \to \tilde{l}_R^+\,\tilde{l}_R^- \to
\tilde{\chi}_1^0\,l^+\,\tilde{\chi}_1^0\,l^-$,
we impose event cuts so that
\begin{itemize}
 \item No jet with $p_T > 20\,\mathrm{GeV}$
 \item Two leptons with $p_T^{l(1,2)} > 50\,\mathrm{GeV}$
       and no other leptons with $p_T^l > 20\,\mathrm{GeV}$
 \item $E_T^{\text{miss}} > 50\,\mathrm{GeV}$
 \item $p_T^{l(1)} + p_T^{l(2)} + E_T^{\text{miss}} >
       200\,\mathrm{GeV}$
 \item The invariant mass of two leptons is outside the region
       $80\,\mathrm{GeV} < m_{ll} < 100\,\mathrm{GeV}$
\end{itemize}
are satisfied.
Note that the last cut is imposed to suppress the SM background.
Since the expected background coming from the SM events is leptonic
decay of $W$ boson and $Z$ boson, we have to consider the $m_{T2}$ distribution
for $WW$, $ZZ$, and $ZW$ production processes.
For the process $p\,p \to W\,W \to l\,\nu\,l\,\nu$, it is obvious
that
\begin{equation}
m_{T2}^{(l)}(0)\leq m_W
\end{equation}
and the process  $p\,p \to Z\,Z \to l\,l\,\nu\,\nu$ is rejected by
the above cut.
For $p\,p \to Z\,W \to l\,l\,l\,\nu$,
this process will contribute to the $m_{T2}$ distribution if one of the
leptons comes from $Z$ boson is missed to be detected. In that case,
\begin{equation}
m_{T2}^{(l)}(0) \le \max\{m_Z, m_W\}=m_Z.
\end{equation}
Therefore, the standard model background does not affect the measurement of
$m_{T2}^{(l)\max}(0)$ if
$(m_{\tilde{l}_R}^2 - m_{\tilde{\chi}_2^0}^2)/m_{\tilde{l}_R} > m_Z$
is satisfied. We checked that these SM background can be negligible
by producing the SM background from the
$ZZ$, $WW$, and $ZW$ processes with the above cuts.

In the model point we are considering here, the maximum value of this $m_{T2}$
variable is given as
\begin{equation}
 \label{eq:M}
 M \equiv m_{T2}^{(l)\max}(0) = 116\,\mathrm{GeV}
\end{equation}
and the result is shown in Figure \ref{fig:q_2}.
Here we illustrate the distribution in the region
$m_{T2}^{(l)}(0) \ge m_Z$, because the SM background is expected to be
negligible only in this region.

The important relation $m_{\tilde{\chi}_2^0} \gg m_{\tilde{\chi}_1^0}$
can be obtained by the calculation as
\begin{equation}
\frac{m_{\tilde{\chi}_2^0}^2}{m_{\tilde{\chi}_1^0}} >
\frac{m_{\tilde{\chi}_2^0}^2}{m_{\tilde{l}_R}} >
\frac{m_{\tilde{\chi}_2^0}^2-m_{\tilde{l}_R}^2}{m_{\tilde{l}_R}} =
\frac{(m_{ll}^{\max})^2}{m_{T2}^{(l)\max}(0)} \sim 2.8\,\mathrm{TeV},
\end{equation}
where we use the measured values of $m_{ll}^{\max}$ and $m_{T2}^{(l)\max}(0)$.

It may be possible to measure the masses of $m_{\tilde{l}_R}$ and
$m_{\chi_1^0}$ by the methods discussed in the papers
\cite{Konar:2009wn,Matchev:2009ad}.
Here we do not discuss this issue further.

\section{Discussion}
Generically, multiple fields may play as the messenger fields.
Actually, any vector-like fields $\Phi_i$ and $\bar\Phi_i$ ($i=1,\cdots, n$)
can be
the messenger fields if they have an interaction with the spurion field
$S$ like $\kappa\Phi_i\bar\Phi_i S$. Here the $F$ component
of $S$ has non-vanishing VEV $F_S$ which breaks the SUSY. 
Naively, they have the same order of the contribution to the SUSY breaking
parameters if the coefficients
$\kappa_i \sim c_i m_{\Phi_i}/\Lambda$ where $\Lambda$ is the cutoff.
Namely, the scale 
$\Lambda_{\Phi}\equiv F_{\Phi_i}/m_{\Phi_i}\sim \alpha_iF_S/\Lambda$ becomes
 independent of $i$ except the coefficients $c_i$ of the
 interactions $m_{\Phi_i}\Phi_i\bar\Phi_i S/\Lambda$. 
Because of the freedom of the $\mathcal{O}(1)$ coefficients, we have
various possibilities
for the sparticle spectrum. Indeed, any gaugino mass spectrum are possible
by choosing the coefficients $c_i$. 
In this paper, we consider an extreme case, in
which one of the coefficients becomes much larger than the others. 
This can often happen when the coefficients are the ratio of the
$\mathcal{O}(1)$ coefficients $c_i = a_i/b_i$, where the
$\mathcal{O}(1)$ coefficients $a_i$ and $b_i$ are determined randomly,
for example, between 0 and 1.
Since it is reasonable to expect that one of the $n$ coefficients of the 
denominator, which is noted as $b_1$, becomes $\mathcal{O}$($1/n$),
the coefficient 
$c_1$ can be $\mathcal{O}$($n$).
If $n \gg 1$, very large coefficient $c_1$ is 
realized. Moreover, if $b_1$ happens to be $\mathcal{O}(1/(10n))$,
which requires
10\% tuning, $c_1 \sim \mathcal{O}(10n)$, and therefore,
the messenger field $\Phi_1$
and $\bar\Phi_1$ can dominate the others. If
$\sum_i^nc_i\sim \mathcal{O}(100)$, then 
since the loop suppression is almost
compensated by the summation of the coefficients, 
the contributions
to the SUSY breaking parameters from the gauge mediation may become of
the same order as those from the direct interactions between the
spurion field and MSSM fields, for example, $Q^\dagger Q S^\dagger S/\Lambda^2$ 
in the K\"ahler potential.
$n$ can be much larger than 1, for example, 
in the anomalous $U(1)$ GUT scenario, there are many
vector-like fields which have non-trivial charges under the standard
gauge group and can be the messenger fields. There are 23 vector-like
pairs in the $SO(10)$ model \cite{Maekawa:2001uk, Maekawa:2001vt},
47 pairs in the $E_6$ model \cite{Maekawa:2002bk},
38 pairs in the simpler $E_6$ model \cite{Maekawa:2003bb}
except the MSSM Higgs doublet pair. 

In this paper, we consider the several cases in which one vector-like
messenger field dominates the others.
Note that although we have chosen
$X+\bar X$ or $Q+\bar Q$ as the messenger field in order to obtain non-vanishing
gaugino masses, other choices become possible if the small but non-zero 
contributions from the other vector-like fields are taken into account.

If there are multiple messengers $\Phi_i$ and $\bar\Phi_i$ ($i=1,\cdots,n$) 
which has the same quantum numbers,
$D$-term contribution of $U(1)_Y$ hypercharge
to the sfermion masses becomes important \cite{Dimopoulos:1996ig}.
This contribution comes from one-loop
Feynman graph and may not be negligible. The explicit formula is given by
\begin{equation}
\Delta m_{\tilde{f}}^2 \sim \frac{1}{2}\sum_{\Phi}
 \left(\frac{\alpha_1}{4\pi}\right)
 Y_{\tilde{f}}Y_\Phi\sum_{i,j}
 \frac{|F_{\Phi_{ij}}|^2-|F_{\Phi_{ji}}|^2}
 {\max\{(m_{\Phi_i})^2,(m_{\Phi_j})^2\}},
\end{equation}
where $F_{\Phi_{ij}}$ is a SUSY breaking mass mixing parameters (so called $B$ parameter) of the messenger fields in the unit in which the mass matrix of the messenger fields is diagonalized as
\begin{equation}
 W_{\text{mess}} = m_{\Phi_i}\Phi_i\bar{\Phi}_i
  + \theta^2F_{\Phi_{ij}}\Phi_i\bar{\Phi}_j.
\end{equation}
If the enhancement factor $c \gg \mathcal{O}(10^2)$,
then such contribution can be negligible.

\section{Summary}
In this paper, we investigated the LHC signatures of the
generalized GMSB models
with the messenger fields which do not respect $SU(5)$ GUT symmetry.
Such a situation can be realized in the anomalous $U(1)$ GUTs
in which the success of the gauge coupling unification can be explained
although the mass spectrum of the vector-like fields do not respect
$SU(5)$ GUT symmetry.
The mass spectrum of sparticles become different from those in the
usual GMSB whose messenger fields respect $SU(5)$ GUT symmetry.
Especially, the gaugino masses do not satisfy the usual GUT relation 
and this feature is very important to distinguish these models
by the LHC measurements.
In principle, any mass pattern for the gaugino masses is possible
in this generalized GMSB scenario.
In this paper, only for simplicity, 
we examined the models with a pair of messenger fields which have
quantum numbers of $X + \bar{X}$ or  $Q + \bar{Q}$
and studied how to obtain the signatures of these models in the LHC.
The gaugino mass relation becomes
$m_{\tilde B}:m_{\tilde W}:m_{\tilde g}\sim 5:6:12$ for the $X + \bar{X}$
messenger model and 
$m_{\tilde B}:m_{\tilde W}:m_{\tilde g}\sim 1/5:6:12$ for the $Q + \bar{Q}$
messenger model.
One of the interesting features of the both models is that the hierarchy
between the colored particle masses and weakly charged particle masses
becomes milder than the usual GMSB models because the messenger fields
have bi-fundamental representation under $SU(3)_C\times SU(2)_L$.
If we catch the scale of the wino mass in the LHC, 
we can roughly check this milder hierarchy by measuring $m_{T2}$ by which
the order of the colored particle masses can be obtained.
In $X + \bar{X}$ messenger scenario, the mass hierarchy between
the bino and wino is very small,
and it leads to the relatively soft $p_T$ distribution of leptons.
On the other hand, the mass difference between the bino and wino is
considerably large in $Q + \bar{Q}$ messenger scenario.
Therefore very large $p_T$ distribution of leptons can be seen at the LHC.

The NLSP of $X + \bar{X}$ messenger
scenario is the stau or neutralino.
If the NLSP is the stau, we can check the mass relation of gauginos
at the low-luminosity stage of the LHC and the deviation from the usual
GUT relation can be obvious.
If the NLSP is the neutralino, we can determine the bino-like and wino-like
neutralino masses by the use of the end-point in the neutralino's leptonic
decay and of the $m_{T2}$ measurement.

In $Q + \bar{Q}$ messenger scenario, the leptonic $m_{T2}$ measurement
is useful because the right-handed slepton remains light.

Since both scenarios predict very characteristic mass spectra,
it is expected that we can distinguish these models from the
models which satisfy the GUT relation.

\section*{Acknowledgments}
We appreciate S.~Asai for his lecture and discussion on the LHC physics.
N.~M. is supported in part by JSPS Grants-in-Aid for Scientific
Research. This research is supported by the Grant-in-Aid for Nagoya
University Global COE Program, ``Quest for Fundamental Principles
in the Universe: from Particles to the Solar System and the Cosmos'',
from the Ministry of Education, Culture, Sports, Science and
Technology of Japan.
\appendix
\section{Two-loop RGE effects for the gaugino mass}
There is a well-known feature in the softly broken SUSY models.
Namely, by looking one-loop RGEs of gauge couplings and gaugino masses,
one can see that their ratio obeys the following RGE.
\begin{equation}
 \label{eq:malpha}
 \frac{d}{d\ln Q}\left(\frac{M_a}{\alpha_a}\right) = 0.
\end{equation}
Then, the gaugino masses in the GMSB model satisfy the relation
\begin{equation}
 \label{eq:gut}
 M_1 : M_2 : M_3 \sim n_1\alpha_1 : n_2\alpha_2 : n_3\alpha_3
\end{equation}
at any renormalization scale.
But it should be noticed that \eqref{eq:malpha} is satisfied only
up to one-loop order, so the actual relation is deviated from
\eqref{eq:gut} by two-loop order correction.

Let us estimate the effect of two-loop order RGE.
The two-loop RGE of gauge couplings and gaugino masses are given by
\begin{equation}
 \frac{d}{d\ln Q}\alpha_a^{-1} = -\frac{b_a}{2\pi}
  - \sum_b\frac{B_{ab}}{8\pi^2}\alpha_b + \frac{c_a}{8\pi^2}\alpha_t
\end{equation}
and
\begin{equation}
 \label{eq:gaugino}
 \frac{d}{d\ln Q}M_a = \frac{b_a}{2\pi}\alpha_aM_a
  + \frac{1}{8\pi^2}\alpha_a
  \left[\sum_bB_{ab}\alpha_b(M_a + M_b)
  + c_a\alpha_t(A - M_a)\right],
\end{equation}
neglecting the Yukawa coupling other than top quark component
$\alpha_t \equiv y_t^2 / 4\pi^2$.
Here, $A$ is defined by $A$-term $A_t \equiv Ay_t$.
If there are no other vector-like particles below the mass scale of
messenger field, $m_{\Phi}$,
the coefficients $b_a$, $c_a$ and $B_{ab}$ are given as
\begin{equation} 
 (b_1,b_2,b_3) = (33/5,1,-3),\qquad
  (c_1,c_2,c_3) = (26/5, 14/5, 18/5)
\end{equation}
and
\begin{equation}
 \begin{pmatrix}
  B_{11} & B_{12} & B_{13} \\
  B_{21} & B_{22} & B_{23} \\ 
  B_{31} & B_{32} & B_{33} 
 \end{pmatrix}
 =
 \begin{pmatrix}
  199 / 25 & 27 / 5 & 88 / 5 \\
  9 / 5 & 25 & 24 \\
  11 / 5 & 9 & 14
 \end{pmatrix}
 .
\end{equation}
Then the deviation of RGE from \eqref{eq:malpha} can be written as
\begin{equation}
 \label{eq:twoloop}
 \frac{d}{d\ln Q}\left(\frac{M_a}{\alpha_a}\right)
  = \sum_b\frac{B_{ab}}{8\pi^2}\alpha_bM_b
  + \frac{c_a}{8\pi^2}\alpha_tA.
\end{equation}

By using one-loop RGE for $A \equiv A_ty_t^{-1}$
\begin{equation}
\frac{d}{d\ln Q}A
 = \frac{3}{\pi}\alpha_tA + \sum_a\frac{d_a}{4\pi}\alpha_aM_a
 + \mathcal{O}(\alpha^2)
\end{equation}
and one-loop part of \eqref{eq:gaugino},
two-loop RGE \eqref{eq:twoloop} can be rewritten as
\begin{equation}
 \label{eq:dm}
 d\left(\frac{M_a}{\alpha_a}\right) = \frac{1}{4\pi}\sum_b
  \frac{B_{ab}}{b_b}dM_b
  + \frac{c_a}{24\pi}
  \left(dA - \sum_b\frac{d_b}{2b_b}dM_b\right) + \mathcal{O}(\alpha^2)
\end{equation}
where
\begin{equation}
 (d_1,d_2,d_3) = (26/15,6,32/3).
\end{equation}
Therefore, integrating \eqref{eq:dm} from messenger mass scale
$m_{\Phi}$ to $m_Z$, we can obtain the gaugino mass formula
at the scale $m_Z$ including the two-loop effect.
Because $A(m_Z) \simeq A_t(m_Z)$, $A(m_{\Phi}) \simeq 0$ and
$M_a(m_Z) \simeq M_a\alpha_a^{-1}(m_{\Phi})\cdot\alpha_a(m_Z)$
at one-loop order, this is expressed as
\begin{align}
 M_a\alpha_a^{-1}(m_Z) &= M_a\alpha_a^{-1}(m_{\Phi})
 + \frac{c_a}{24\pi}A_t(m_Z) \notag \\
 &\qquad - \frac{1}{4\pi}\sum_b
  \left(\frac{B_{ab}}{b_b} - \frac{c_ad_b}{12b_b}\right)
  M_b\alpha_b^{-1}(m_{\Phi})\Delta\alpha_b,
\end{align}
where $\Delta\alpha_a \equiv \alpha_a(m_{\Phi}) - \alpha_a(m_Z)$.
If we write $\Lambda_{\Phi} \equiv F_{\Phi} / m_{\Phi}$ and
$r_t \equiv -4\pi A_t(m_Z)/\Lambda_{\Phi}$,
we finally obtain a following result.
\begin{equation}
 M_a\alpha_a^{-1}(m_Z) =
  \left[n_a - \frac{c_a}{24\pi}r_t
   - \sum_b\frac{n_b}{4\pi}
  \left(\frac{B_{ab}}{b_b} - \frac{c_ad_b}{12b_b}\right)
  \Delta\alpha_b\right]\frac{\Lambda_{\Phi}}{4\pi}.
\end{equation}
Since the order of $r_t$ and $\Delta\alpha_a$ are $\mathcal{O}(0.1)$
at most, two-loop contribution cannot become so large in the
typical case, although it can becomes important if $n_a \ll 1$.
\numberwithin{equation}{section}
\renewcommand{\theequation}{\Alph{section}.\arabic{equation}}
\end{document}